\documentclass[%
aip,
jcp,
 amsmath,amssymb,
reprint,%
]{revtex4-1}
\usepackage{graphicx}
\usepackage{dcolumn}
\usepackage{bm}
\usepackage{placeins}
\usepackage{array,mathtools,amssymb,booktabs}
\AtBeginDocument{
\heavyrulewidth=.08em
\lightrulewidth=.05em
\cmidrulewidth=.03em
\belowrulesep=.65ex
\belowbottomsep=0pt
\aboverulesep=.4ex
\abovetopsep=0pt
\cmidrulesep=\doublerulesep
\cmidrulekern=.5em
\defaultaddspace=.5em
}

\usepackage{chngcntr}
\usepackage{float}

\usepackage[dvipsnames]{xcolor}
\usepackage{tikz}
\usetikzlibrary{shapes}
\definecolor{mBlue}{RGB}{0,148,200}
\definecolor{mOrange}{RGB}{255,140,0}
\definecolor{mRed}{RGB}{200,0,0}
\definecolor{mGreen}{RGB}{68,85,37}
\DeclareRobustCommand{\hc}{\raisebox{0.4pt}{\tikz{\node[draw,scale=0.48,circle,fill=mBlue](){};}}}
\DeclareRobustCommand{\Tc}{\raisebox{0.4pt}{\tikz{\node[draw,scale=0.48,circle,fill=mRed](){};}}}
\DeclareRobustCommand{\darkgreenstar}{\raisebox{0.35pt}{\tikz{\node[draw,scale=0.3,star, star points=5, star point ratio=2.25, fill=mGreen](){};}}}
\DeclareRobustCommand{\circles}{\raisebox{0.4pt}{\tikz{\node[draw,scale=0.48,circle,fill=mBlue](){};}}}
\DeclareRobustCommand{\triangle}{\raisebox{0.35pt}{\tikz{\node[draw,scale=0.3, minimum height=1em,regular polygon,regular polygon sides=3,fill=mOrange](){};}}}
\DeclareRobustCommand{\diamond}{\raisebox{0.2pt}{\tikz{\node[draw,scale=0.25, minimum height=3em,diamond,fill=mRed](){};}}}

\begin{document}

\preprint{--}

\title{Model for disordered proteins with strongly sequence-dependent liquid phase behavior}
%
\author{Antonia Statt}
\thanks{H. Casademunt and A. Statt contributed equally to this work.}
\affiliation{Department of Chemical and Biological Engineering, Princeton University, Princeton, NJ 08544}
\affiliation{Present address: Materials Science and Engineering, Grainger College of Engineering, University of Illinois, Urbana-Champaign, IL,61801 } 

\author{Helena Casademunt}%
\thanks{H. Casademunt and A. Statt contributed equally to this work.}
\affiliation{Department of Physics, Princeton University, Princeton, NJ 08544}%
\affiliation{Present address: Department of Physics, Harvard University, Cambridge, MA 02138}


\author{Clifford P. Brangwynne}
\affiliation{Department of Chemical and Biological Engineering, Princeton University, Princeton, NJ 08544}
\affiliation{Howard Hughes Medical Institute, Chevy Chase, MD 20815}

\author{Athanassios Z. Panagiotopoulos}
\email[email:]{azp@princeton.edu}
\affiliation{Department of Chemical and Biological Engineering, Princeton University, Princeton, NJ 08544}
\date{\today}

\begin{abstract}
Phase separation of intrinsically disordered proteins is important for the formation of membraneless organelles, or biomolecular condensates, which play key roles in the regulation of biochemical processes within cells. In this work, we investigated the phase separation of different sequences of a coarse-grained model for intrinsically disordered proteins and discovered a surprisingly rich phase behavior. We studied both the fraction of total hydrophobic parts and the distribution of hydrophobic parts. Not surprisingly, sequences with larger hydrophobic fractions showed  conventional liquid-liquid phase separation. The location of the critical point was systematically influenced by the terminal beads of the sequence, due to changes in interfacial composition and tension. For sequences with lower hydrophobicity, we observed not only conventional liquid-liquid phase separation, but also reentrant phase behavior, in which the liquid phase density decreases at lower temperatures. For some sequences, we observed formation of open phases consisting of aggregates, rather than a normal liquid. These aggregates had overall lower densities than the conventional liquid phases, and exhibited complex geometries with large interconnected string-like or membrane-like clusters. Our findings suggest that minor alterations in the ordering of residues may lead to large changes in the phase behavior of the protein, a fact of significant potential relevance for biology. 
\end{abstract}

\pacs{Valid PACS appear here}
\maketitle
\section{Introduction}

Liquid-liquid phase separation of intrinsically disordered proteins (IDPs) in cells is known to be crucial for a number of biological functions.~\cite{Berry2018,Shin2018,Shin2017,Banani2017} Membraneless organelles (such as the nucleolus, stress granules, P-bodies, and many more~\cite{Zhu2015}) partition cellular components into distinct regions, which play an important role in regulating biochemical processes.~\cite{Berry2015}

IDPs adopt many different chain conformations, much like synthetic polymers. This directly contributes to phase separation properties.~\cite{Uversky2017} The unfolded conformational states are hypothesized to enable the formation of  transient interaction networks,~\cite{Zhou2018} which make phase separation possible at far lower concentrations than for folded proteins.~\cite{Wei2017} 

The precise importance of protein disorder remains unclear. Here, we address the question of how much the sequence of the protein matters for phase separation.
In order to better understand the biological relevance of IDPs and their phase separation, we attempt to make a connection between their sequence, conformational distributions, and resulting phase behavior.~\cite{Wright2015}

The substructure and properties of aggregates of IDPs are especially relevant to their biological function. Certain membraneless organelles (like those formed by FUS, an RNA-binding protein~\cite{Burke2015}) are known to be liquid-like, while others (e.g. TDP-43, a DNA-binding protein involved in splicing regulation~\cite{Molliex2015}) have a more gel-like structure. For certain IDPs, changes in sequence-dependent phase behavior have been shown to drive pathological aggregation. This is the case for mutations in FUS and TDP-43 that are associated with amyotropic lateral sclerosis (ALS).~\cite{Babinchak2019}
Aggregate formation is commonly observed and well characterized in block copolymers,~\cite{Koch2015,*Floriano1999,Li2019,Posocco2010,Dolgov2018} where self-assembly is driven by the microphase separation of the different blocks, but it is less well understood for proteins. 

Reentrant phase behavior, where the concentration of the dense phase first increases, reaches a maximum, and then decreases again, is proposed to be an important regulatory mechanism for dissolving membraneless organelles,~\cite{Milin2018} as well as for the  formation of dynamic droplet substructures or vacuoles.~\cite{Banerjee2017} This type of phase behavior is found in many different systems, such as patchy particles,~\cite{Espinosa2019} network forming systems,~\cite{Russo2011,Zilman2002} 
and proteins.~\cite{Tempel1996,Banerjee2017,Milin2018,Zhang2008,Zhang2010,Moller2014,Jordan2014} In proteins, it can be driven by temperature,~\cite{Tempel1996} RNA concentration,~\cite{Banerjee2017,Milin2018} or salt/ion concentration.~\cite{Jordan2014,Zhang2008,Zhang2010}

To investigate the phase behavior of IDPs, both theoretical calculations~\cite{Mccarty2019,Lee2008,Sawle2015} and molecular dynamics simulations on the atomistic~\cite{Das2013,Das2018,Mao2010,Vitalis2009,Wei2017,Zerze2015} and the coarse-grained level~\cite{Dignon2019,Dignon2018,Mccarty2019,Qin2016} have recently been performed. Proteins can also be modeled with lattice models,~\cite{OToole1992} or as simple patchy particles or multi-component mixtures of patchy particles.~\cite{Nguemaha2018,Sarangapani2015,Liu2007,Ghosh2019} Recently, there has been significant progress in using the sticker and spacer model,~\cite{Harmon2017,Wang2018} which preserves the polymeric nature of proteins. Field theory based methods~\cite{Lin2016,Lin2017,Mccarty2019} have been used to show the effects of charge patterning on the phase behavior of an IDP.

In this work, we focus on the sequence dependence of phase and aggregation behavior rather than the bulk self-assembly of chains, which has been studied extensively for various architectures of synthetic polymers,~\cite{Matsen2012,Zhang2017,Bates2017,Levine2016} including multiblock copolymers~\cite{Bates2012,Wu2004} and tapered blocks.~\cite{Pakula1996} Aggregation behavior has also been extensively studied for dilute systems in solution, specifically with di- and tri-blocks~\cite{Dolgov2018,Li2019} and multiblocks,~\cite{Gindy2008} where the focus was mainly to describe finite-size aggregates like micelles and vesicles, and gelation.~\cite{Hugouvieux2009,*Hugouvieux2011}


In the present work, we use a simplified model of an IDP, where each section of the protein has either favorable or unfavorable interactions with the surrounding solvent and itself, which we call hydrophobic or hydrophilic hereafter. Because of the computational efficiency of the model, we are able to systematically study the influence of both the overall level of hydrophobicity and the distribution of hydrophobic/hydrophilic regions on liquid-liquid phase separation, as well as the character of the aggregates that form. When the sequence has a substantial amount of both hydrophobic and hydrophilic beads, the large number of possible different sequences allows us to investigate the influence of the bead distribution on the phase behavior. We note that none of the sequences studied in this work corresponds to a specific protein. Instead, we aim at uncovering and understanding systematic trends in the phase behavior of these model disordered proteins.

In the following, we first describe the model and simulation details in section~\ref{sec:model}, and 
then investigate the influence of hydrophobicity on the phase separation in section~\ref{sec:phasesep}, as well as the role of the end of the chain in section~\ref{sec:phasesep2}. We study the effect of the distribution of hydrophobic parts in section~\ref{sec:ft06distribution}, where we observe reentrant phase behavior and large-scale aggregation. 
We then investigate a number of previously proposed order parameters in section~\ref{sec:orderparameter}. Finally, we conclude with discussion and outlook in section~\ref{sec:outlook}.

\section{Model and Methods \label{sec:model}}

In this work, we study the phase behavior of a simplified model for IDPs using classical molecular dynamics (MD)
simulations. We investigate the influence of the distribution of hydrophobic/hydrophilic regions and the degree of hydrophobicity on the resulting phase behavior. 
Thus, we only use two types of regions (``beads" in the model), namely hydrophobic and hydrophilic.
For computational efficiency, we use an implicit-solvent model, so the vapor and liquid phases correspond to the dilute and condensed liquid phases of an IDP solution. 
Each chain consists of $M=20$ bonded beads of mass $m$ each. 
Because of the coarse-grained nature of this model, each bead corresponds to multiple amino-acids in a protein. The length $M=20$ ensures that the chains are not entangled.  

Inspired by surfactant models, we name the  hydrophilic beads H, and the hydrophobic beads T. 
The hydrophobic, attractive T beads interact through the Lennard-Jones (LJ) potential
\begin{align}
    U_\text{LJ}(r) = 4 \epsilon \left[ \left(\frac{\sigma}{r}\right)^{12} - \left(\frac{\sigma}{r}\right)^{6}\right] \quad,
\end{align}
where $r$ is the distance between two beads, $\epsilon$ is the energy well depth and $\sigma$ determines the interaction range. For computational efficiency, we applied a smoothing function to gradually decrease both the force and potential to zero at a cutoff of $r=3\sigma$. The functional form can be found in the SM. The pair interaction of hydrophilic H beads was modeled with a purely repulsive Weeks-Chandler-Anderson (WCA) potential~\cite{Weeks1971}
\begin{align}
    U_\text{WCA}(r) =
    \begin{cases} 
    U_\text{LJ}(r) +\epsilon & r< 2^{1/6}{\sigma}\\
    0 & r\geq 2^{1/6}{\sigma}\\
    \end{cases}\quad .
\end{align}
Cross-interactions between hydrophilic and hydrophobic beads were also described by the WCA potential. The total fraction of attractive, hydrophobic beads along the chain is denoted by $f_T$.  Bonds between subsequent beads in the chain are described by the FENE potential
\begin{align}
    U_\text{b}(r) = -\frac{KR_0^2}{2}\ln{\left[ 1 - \left(\frac{r}{R_0}\right)^2\right]} + U_\text{WCA}(r) \quad,
\end{align}
where $R_0=1.5\sigma$ is the maximum extension of a bond and $K=30\epsilon/\sigma^2$ is the spring constant. 

All simulations were performed using the HOOMD-blue (version 2.6.0) simulation package~\cite{Glaser2015,*Anderson2008} on graphics processing units. The equations of motion were integrated using the velocity-Verlet algoritm with a timestep of $0.005\tau$, where $\tau = \sqrt{m\sigma^2/\epsilon}$ is the unit of time. A weakly coupled Langevin thermostat with a friction constant of $0.1m/\tau$ was employed in the $NVT$ simulations to keep temperature constant. 
In the case of $NpT$ simulations, a MTK barostat-thermostat~\cite{Martyna1994} with coupling constants $\tau=0.5$ and $\tau_P=0.5$ was used. 
In the following, $\epsilon$ is used as the energy unit,  $\sigma$ as the unit of length, and the mass $m$ of a single bead as the unit of mass. 

To obtain the coexistence properties we used the established direct coexistence method.~\cite{Rowlinson1982} Coexisting dense and dilute phases were simulated in an elongated box with dimensions $L_x \times L_y \times L_z$, where $L_z>L_x=L_y$. The two interfaces present in the simulation were oriented perpendicular to $z$ to minimize surface energy. By recording density profiles along $z$, the coexistence densities were estimated from the density of the bulk regions sufficiently far away from the interfaces. 

To check for finite size effects, we systematically varied both the cross-sectional area $A=L_x^2$ and the length of the simulation box ranging from $L_x=10\sigma$ to $50\sigma$ and $L_z=3L_x$ to $6 L_x$ for selected sequences at different temperatures. We found that a box size of $L_x=L=30\sigma$ and $L_z=5L$ with  $N=1000$ chains of length $M=20$ is sufficient to obtain reliable values for the coexistence densities. We excluded any simulations where either bulk phase occupied less than $10\sigma$ in the $z$ direction and repeated them in a larger box of size $50\sigma\times 50\sigma \times 250\sigma$ containing $N=4000$ chains, increasing the volume of both bulk phases. 

All simulations were run on a single GPU (NVIDIA P100 or NVIDIA GeForce GTX 1080) for at least $25,000\tau$ for equilibration and then another $50,000\tau$ for measuring the density histograms from which the coexistence densities were obtained.  The critical points $(\rho_c,T_c)$ were estimated using the universal scaling of the coexistence densities near the critical point and the law of rectilinear diameters
\begin{align}
   \rho_L - \rho_v &= \Delta \rho_0 (1- T/T_c)^\beta \quad , \label{eq:univ} \\
    \frac{\rho_L + \rho_v}{2} &= \rho_c + A (T_c -T) \quad, \label{eq:rect}
\end{align}
where $\beta \approx 0.325$ is the three-dimensional Ising model critical exponent.~\cite{Rowlinson1982} $A$ and $\Delta \rho_0$ are system specific fitting parameters. Because Eq.~(\ref{eq:univ}) is only valid close to the critical point, we only fitted coexistence densities up to approximately $30\%$ below the critical point. Any simulations close to the critical temperature where the standard deviations of the coexistence densities were larger than the difference between the two densities were also excluded. 

To estimate the statistical uncertainties in the critical points we used the statistical error in the coexistence densities. Each run was divided into ten equal parts and the coexistence densities were determined for each of the parts independently. Then, we fitted Eqs.~(\ref{eq:univ})--(\ref{eq:rect}) 300 times with a randomly selected coexistence density out of the ten parts for each measured point.~\cite{Silmore2017}

\section{Results and Discussion}
\subsection{Influence of the fraction of hydrophobic beads on the phase behavior \label{sec:phasesep}}

\begin{figure}
    \centering
    \includegraphics[width=\columnwidth]{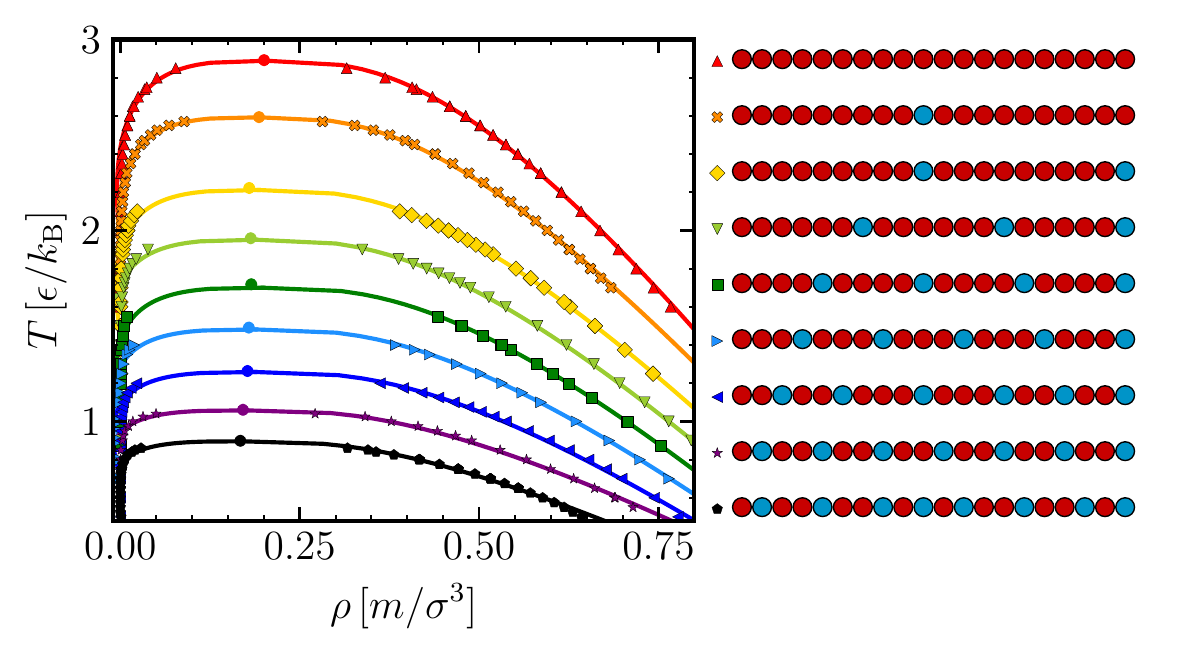}
    \caption{Coexistence densities of the dilute and dense phases for regular sequences with varying $f_T$, from $f_T=1.0$ (red) to $f_T=0.6$ (black), as indicated. The locations of the critical points are shown by circles and the lines show the fit of Eqs.~(\ref{eq:univ})--(\ref{eq:rect}) to the upper part of the phase envelope. The legend shows the corresponding sequences (sorted from high to low $T_c$) with filled red circles for T beads (\Tc) and filled blue circles for H beads (\hc). Statistical uncertainties are smaller than symbol size.}
    \label{fig:phase_envelopes}
\end{figure}

We first determined the phase diagrams for a number of regular sequences with different degrees of hydrophobicity, starting from the fully hydrophobic chain. We systematically varied the fraction of hydrophobic beads from $f_T=1$ to $f_T=0.6$ by adding repulsive beads evenly distributed along the chain. The measured phase diagrams of selected sequences are shown in Fig.~\ref{fig:phase_envelopes}.  In this and subsequent figures, sequences are depicted with hydrophobic, attractive T beads in red (\Tc) and hydrophilic, repulsive H beads in blue (\hc). All sequences studied and their respective critical points can be found in table~\ref{tab:crit} in the SM. 

As shown in Fig. 1, the critical temperature decreases with decreasing fraction $f_T$ and the phase envelopes become flatter and narrower, as expected from phase behavior of long chains.~\cite{Silmore2017} Note that all simulated chains in this work are of the same length $M=20$. The shifts in phase envelope and critical points are purely due to the fraction of attractive beads and their distribution along the chain.

\begin{figure}
    \centering
    \includegraphics[width=\columnwidth]{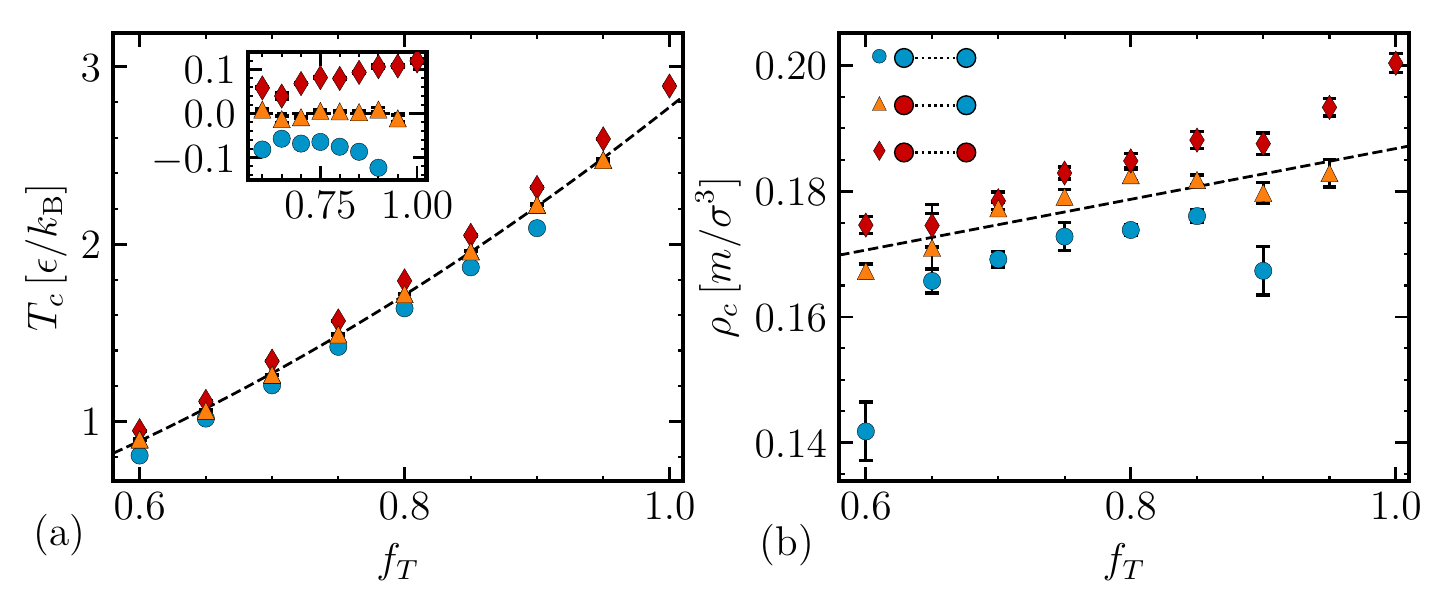}
    \caption{Scaling of the critical temperature $T_c$ (a) and critical density $\rho_c$ (b) with $f_T$. As shown in the legend, red diamonds (\diamond) indicate sequences where both terminal beads are T, blue circles (\circles) mark sequences with two H ends, and orange triangles (\triangle) indicate the sequences with one H and one T end. The dashed line in (a) indicates a quadratic fit and a linear fit to the data in (b). The inset in (a) shows the deviation of $T_c$ from the quadratic fit.}
    \label{fig:crit_scaling}
\end{figure}

As expected from Flory-Huggins scaling,~\cite{Flory1942,*Huggins1942} the critical temperature $T_c$ shows a quadratic dependency on the fraction of hydrophobic, attractive beads $f_T$ as shown in Fig.~\ref{fig:crit_scaling}(a). A systematic decrease with decreasing $f_T$ can be observed in the critical densities $\rho_c$ as well (Fig.~\ref{fig:crit_scaling}(b)). 

Increasing the fraction of hydrophobic, attractive beads in our model is similar to increasing a protein's number of stickers, representing folded binding domains.~\cite{Holehouse2018} Studies have shown that performing mutations at sticker sites, effectively rendering them non-functioning, reduces the propensity of a protein to phase separate,~\cite{Bracha2018} which is consistent with our observations. 

\subsection{Influence of the terminal bead type on the location of the critical point \label{sec:phasesep2}}

The findings above show that the fraction of hydrophobic, attractive beads is important for defining the phase boundary, but it is unclear if the precise sequence of beads matters.
For each $f_T\leq0.9$, we computed the phase boundaries and critical point for (1) a sequence with two attractive, hydrophobic terminal beads, (2) a sequence with two repulsive, hydrophilic terminal beads, and (3) a sequence with mixed terminal beads. All sequences can be found in table~\ref{tab:crit} in the SM.
 
Attractive ends shifted both $T_c$ and $\rho_c$ up, whereas repulsive ends resulted in a lowered  $T_c$ and $\rho_c$, with the mixed end sequences in between. This effect appeared to be general for all investigated sequences up to $f_T=0.6$, as shown in Fig.~\ref{fig:crit_scaling}.
 
As previously suggested, the radius of gyration $R_g$ of a single chain measured at a fixed temperature could be used as a predictor of the critical point.~\cite{Dignon2018,Lin2017,Lin2017_2} It is especially useful because the radius of gyration is an experimentally accessible quantity.~\cite{Hofmann2012,Riback2017} For the model investigated here, we indeed found that both critical temperature and density scale roughly linearly with $R_g$, as reported in Fig.~\ref{fig:TRgThetaScaling} in the SM. However, $R_g$ did not capture the influence of the terminal bead type, suggesting that this effect results from some other mechanism.

\begin{figure}
    \centering
    \includegraphics[width=0.45\textwidth]{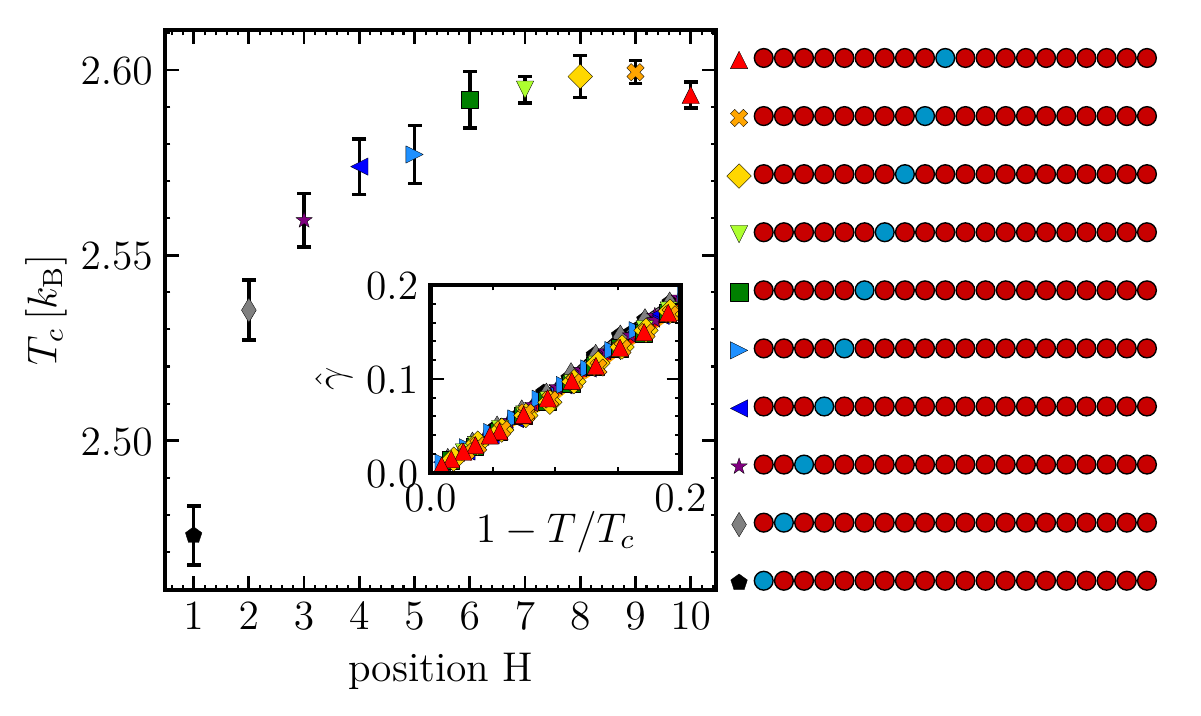}
    \caption{Scaling of the critical temperature as a function of the position of the repulsive H (\hc) bead from the end of the chain to the middle. The inset shows the scaling of $\hat \gamma$ with the distance $1-T/T_c$ to the critical point. }
    \label{fig:scalingT19gamma}
\end{figure}

To systematically study the effect of terminal bead type, we simulated a series of chains with only one repulsive bead ($f_T=0.95$) and varied its position along the chain from the middle to the end. The phase diagrams and scaling of the critical point are shown in Fig.~\ref{fig:T19phase} in the SM. For sequences where the repulsive bead was about 6 positions away from the end or more, almost no difference in critical point was measured. For sequences where the repulsive bead was near the end of the chain, we observed that the critical point decreased as the bead moved closer to the end. As displayed in Fig.~\ref{fig:scalingT19gamma}, the decrease of the critical point was approx. $5\%$ when moving the repulsive, hydrophilic H bead from the middle to the end to the chain, and approx. $3\%$ when moving the H bead to the second outermost position of the chain.

The type of terminal bead had a systematic influence on the interfacial composition, as shown in Fig.~\ref{fig:T19interface} in the SM. A slight excess of chain ends at the dilute-dense interface is expected because of entropic effects, even for the homopolymer case.~\cite{Helfand1989} We observed an enhancement in the concentration of both end beads and repulsive beads in the interfacial region for sequences where the repulsive H bead was near the end of the chain. The interface composition influences the value of the interfacial tension, which scales according to
\begin{align}
    \gamma = \gamma_0 (1-T/T_c)^\mu \label{eq:gamma}
\end{align}
with the distance to the critical point, where $\mu \approx 1.26$, the relevant exponent for the 3-dimensional Ising universality class.~\cite{Widom2013} Because the critical point and the interfacial tension are connected, we speculate that any changes in interfacial composition also lead to a change in the critical point.  

Due to computational limitations, we have not measured the interfacial tension directly in this work, but instead report an estimated interfacial tension $\hat \gamma = k_\text{B}T/(2\pi w^2) \ln L$ from the interface width $w$, as shown in the inset of Fig.~\ref{fig:scalingT19gamma}. As expected from Eq. (\ref{eq:gamma}), we observed a collapse of $\hat \gamma$ onto the same curve for the different sequences, when plotted against $1-T/T_c$. The investigation of the precise relationship between sequence, critical point and interfacial tension is left for future work.

Thus, these findings show that the exact sequence can have a significant effect on the phase boundary and signal the importance of the terminal bead, a result which to our knowledge has not been reported before.
It would be interesting to study experimentally if mutations at the ends of an IDP have a more pronounced effect on the phase behavior than mutations in the middle of the chain. To our knowledge, the effect of the relative position of a sticker site on the phase behavior of a protein has not been experimentally probed.

\subsection{Influence of the distribution of attractive beads on the phase behavior for  $\bf f_T=0.6$ ~\label{sec:ft06distribution}}

In contrast to the sequences with higher fraction of attractive beads, the chains with $f_T = 0.6$, the lowest value studied, displayed not only conventional phase separation between a dilute and a dense phase, but also reentrant phase behavior. This behavior is characterized by a density of the condensed phase that first increases, reaches a maximum, and then decreases as temperature decreases (see Fig.~\ref{fig:T12reentrant}). We also found sequences that do not seem to form conventional liquid phases at all, but form large-scale aggregates instead. In the following, we describe results for each of these three different cases.

\subsubsection{ Phase separation and reentrant phase behavior
\label{sec:reentrant}}

We have found some sequences with $f_T = 0.6$ (shown in Fig.~\ref{fig:T12liquid} in the SM) that exhibit conventional phase separation into a dilute and a dense phase, following Eqs. (4)--(5). By conventional phase separation we mean that it involves a first-order phase transition with a discontinuous change in density. 
Because we did not exhaustively investigate the space of possible sequences, we expect that there are more sequences with conventional phase separation. 
The glass transition for the purely attractive homopolymer is $T_g\approx0.4\epsilon/k_\text{B}$,~\cite{Jain2004} and we therefore did not attempt to simulate temperatures below $0.45 \epsilon/k_\text{B}$.

\begin{figure}
    \centering
    \includegraphics[width=0.49\textwidth]{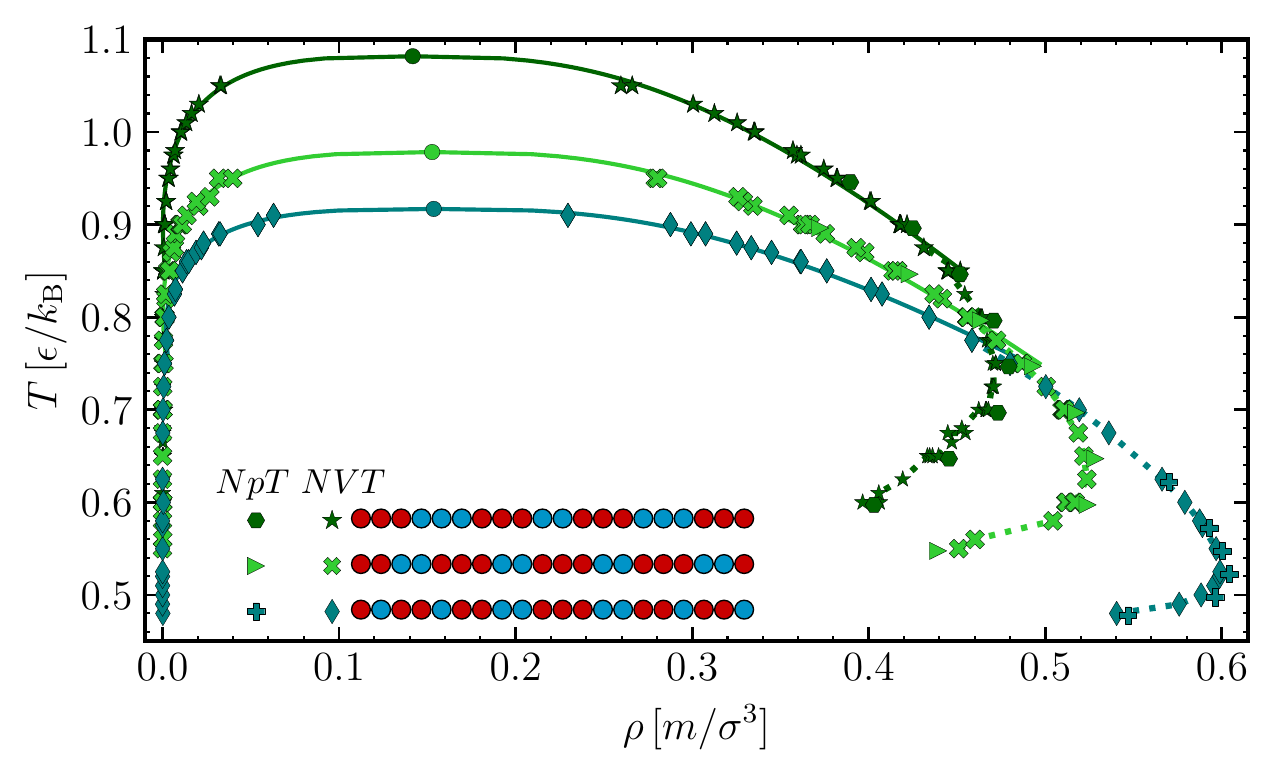}
        \caption{Coexistence densities for selected sequences with $f_T=0.6$ which show reentrant behavior. The coexistence densities were measured in $NVT$ and $NpT$ ensembles as indicated. The locations of the critical points are marked by circles and the statistical uncertanties are smaller than the symbol size. The solid lines show the fit of Eqs.~(\ref{eq:univ})--(\ref{eq:rect}) to the upper part of the phase envelope determined in $NVT$ conditions. The dashed lines are guides to the eye only.}
    \label{fig:T12reentrant}
\end{figure}

Some sequences with $f_T=0.6$ showed reentrant phase behavior, where the density of the condensed phase first increased, reached a maximum, and then decreased with decreasing temperature. Examples of the phase envelopes are displayed in Fig.~\ref{fig:T12reentrant}; the rest of the sequences can be found in Fig.~\ref{fig:T12liquid} in the SM. 
First, a conventional phase separation into a dense and a dilute phase occurred as the system was cooled down. This part if the phase envelope can be fitted by Eqs.~(\ref{eq:univ})--(\ref{eq:rect}). Upon further cooling, the dense phase developed sub-structures. We observed small H and T rich regions, with large voids in between them. This microphase separation of the condensed liquid resulted in a lower overall density of the dense phase.

\begin{figure}
    \centering
    \includegraphics[width=0.45\textwidth]{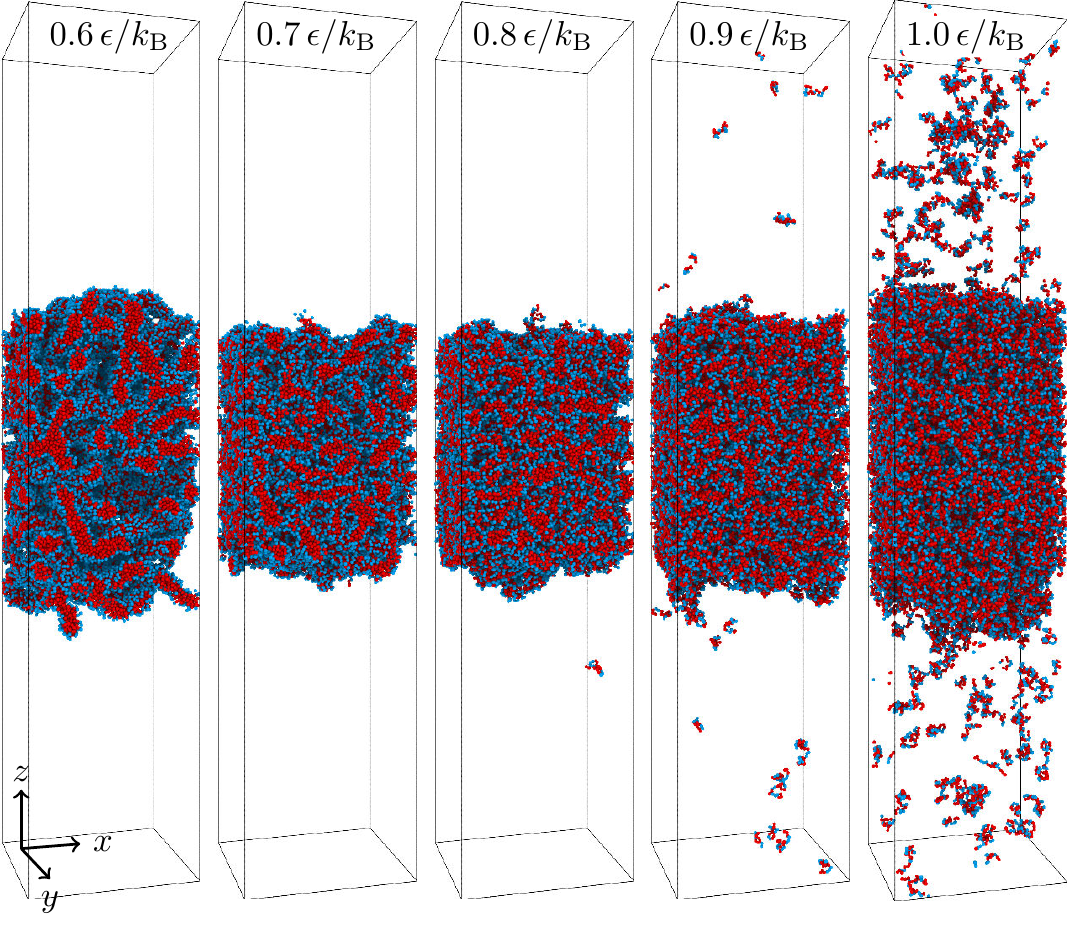}
    \caption{Typical configurations of the sequence $\text{T}_3\text{H}_3\text{T}_3\text{H}_2\text{T}_3\text{H}_3\text{T}_3$ (\Tc\Tc\Tc\hc\hc\hc\Tc\Tc\Tc\hc\hc\Tc\Tc\Tc\hc\hc\hc\Tc\Tc\Tc) at different temperatures. The snapshots were taken from the direct coexistence $NVT$ simulations in a box of size $50\sigma \times 50\sigma \times 250\sigma$ with $N=4000$ chains. Hydrophobic, attractive T beads are shown in red (\Tc) and hydophilic H beads are shown in blue (\hc). The critical point is $T_c = (1.109\pm 0.006) \varepsilon/k_B$, $\rho_c=(0.146 \pm 0.002) m/\sigma^3$ and the phase envelope (\darkgreenstar) of this sequence is shown in Fig.~\ref{fig:T12reentrant}. Snapshots were generated using OVITO.~\cite{ovito} }
    \label{fig:snapshots_reentrant}
\end{figure}

For even colder temperatures, we found formation of large-scale aggregates. A typical example for the sequence $\text{T}_3\text{H}_3\text{T}_3\text{H}_2\text{T}_3\text{H}_3\text{T}_3$ 
is shown in Fig.~\ref{fig:snapshots_reentrant}, and more examples at different temperatures can be found in Fig~\ref{fig:snapshots_strings} in the SM. These large-scale aggregates typically had fairly low densities compared to the liquid densities observed for conventional phase separation into a disordered condensed phase. The structure and properties of the aggregates are discussed in the next section~\ref{sec:aggregate}.
 
We confirmed the results of the $NVT$ direct coexistence measurements with $NpT$ simulations of the dense phase (shown in Fig.~\ref{fig:T12reentrant}). For the $NpT$ simulations, we estimated $p\approx0$ from the ideal gas law $p \approx (\rho_v/Mm)k_\text{B}T$. Accordingly, we only performed $NpT$ simulations for low temperatures, where $\rho_v\approx 0$. 

For the reported temperature range, we did not observe a difference in the structure or density of the phases between $NVT$ and $NpT$ or between different independent simulation runs. For sufficiently cold temperatures, we were unable to equilibrate the large-scale aggregates reliably and excluded the data. 

The critical temperatures and densities of the reentrant sequences in Fig.~\ref{fig:T12reentrant} were comparable to those of the sequences in Fig.~\ref{fig:phase_envelopes}. Even though the reentrant onset in the liquid branch occurred at different temperatures for each sequence, the relative distance $T/T_c$ to the critical temperature is similar for all of them, roughly 60-70\% below $T_c$, suggesting a common underlying mechanism. However, the highest density value for the liquid branch varies greatly between $0.45$ and $0.6 m/\sigma^3$ as visible in Fig.~\ref{fig:T12reentrant}.

We observed limitations with respect to our ability to equilibrate the systems at  low temperatures: (1) near the glass transition, the dynamics slowed down drastically, (2) the results depended on the initial configuration and ensuring that large-scale structures were properly equilibrated became increasingly more difficult, and (3) for some large-scale aggregates it was not clear how to unambiguously define a bulk density. Therefore, we limited each reported phase diagram to the temperature range where the mentioned limitations were not severe. 

The sequences that exhibited reentrant phase behavior have an overall more ``blocky'' distribution of hydrophilic and hydrophobic sections, with the longest section being three beads long. This suggests that ``blockiness'' plays an important role in the ability to form structured liquids at low temperature, in agreement with Nott et al.,~\cite{Nott2015} who showed the relevance of blocky patterned electrostatic interactions to phase separation. We found all three possible combinations of terminal end beads amongst the sequences as well as different terminal block lengths. Because we investigated a limited subset of possible sequences, future work will be needed to solidify the systematic connection between sequences and phase behavior.

The driving mechanism for reentrant phase behavior is the competition between self-assembly of large-scale ordered structures and condensation into a conventional dense liquid, similar to what has been observed in patchy particles~\cite{Espinosa2019} and network-forming fluids.~\cite{Russo2011} We observed the formation of large-scale microphase separated structures at low temperatures, which led to a lower density of the condensed phase. It is therefore reasonable to assume that the temperature at which we observe reentrant phase behavior will be connected to the order-to-disorder transition temperature~\cite{Pakula1996,Gindy2008} of the sequence, which determines the temperature dependence of the microphase separation.


\FloatBarrier
\subsubsection{Formation of large-scale aggregates and their structure 
\label{sec:aggregate}}

\begin{figure}
    \centering
    \includegraphics[width=0.5\textwidth]{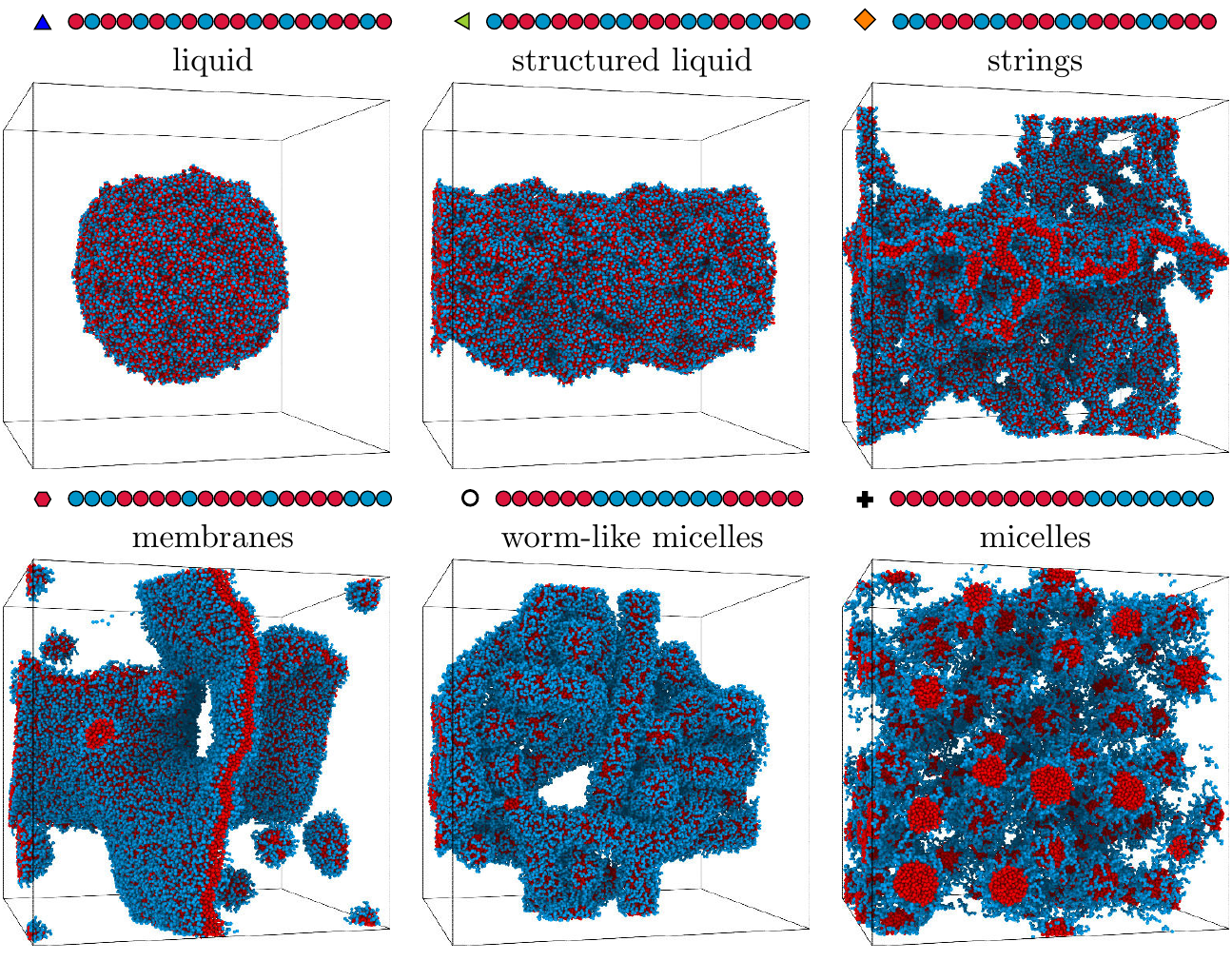} 
    \caption{Typical configurations at $\rho_\text{tot}=0.1m/\sigma^3$ and $T=0.65\epsilon/k_\text{B}$ of selected sequences with $f_T=0.6$. See main text for identification criteria. The volume of each system was $100\sigma^3$. Snapshots were generated using OVITO.~\cite{ovito}}
    \label{fig:snapshots_other}
\end{figure}

In addition to conventional dense-dilute phase separation and reentrant phase behavior, we also found some sequences which only formed large-scale aggregates. These sequences are listed in table~\ref{tab:nocrit} in the SM. In contrast to the previously discussed sequences in section~\ref{sec:reentrant}, the sequences discussed here did not show dense-dilute phase coexistence at any given temperature. Instead, the large-scale aggregates fell apart into smaller isolated aggregates when the temperature was increased sufficiently. 

The structure of the  large-scale aggregates  varied strongly with sequence. We observed fibril-like or string-like clusters that were interconnected, as displayed in Fig.~\ref{fig:snapshots_other}. This type of aggregate is characterized by many large voids. We also found membrane-like structures for some sequences (also in Fig.~\ref{fig:snapshots_other}), including empty vesicles, flat membranes, and layered, onion-like structures, often observed in the same simulation. Similar aggregate structures are reported in the literature for multi-block polymers.~\cite{Wu2009,Li2019,Dolgov2018,Kuldova2013,Gindy2008,Hugouvieux2009,*Hugouvieux2011}

In addition to the large-scale fibril-like aggregates and membrane-like aggregates, we also found the expected finite sized aggregates from multi-block polymer literature: worm-like interconnected micelles for the tri-block chain $\text{T}_6\text{H}_8\text{T}_6$, 
and spherical micelles for $\text{H}_4\text{T}_{12}\text{H}_4$ 
and the block-copolymer $\text{T}_{12}\text{H}_8$. 
Some examples for those morphologies are shown in Fig.~\ref{fig:snapshots_other}. None of the measured short range structural properties of the chains in those aggregates and the liquid configurations of the previous sections were different enough to distinguish between the different behaviors, as shown in section~\ref{sec:structureliquid} in the SM.  

\begin{figure}
    \centering
    \includegraphics[width=0.45\textwidth]{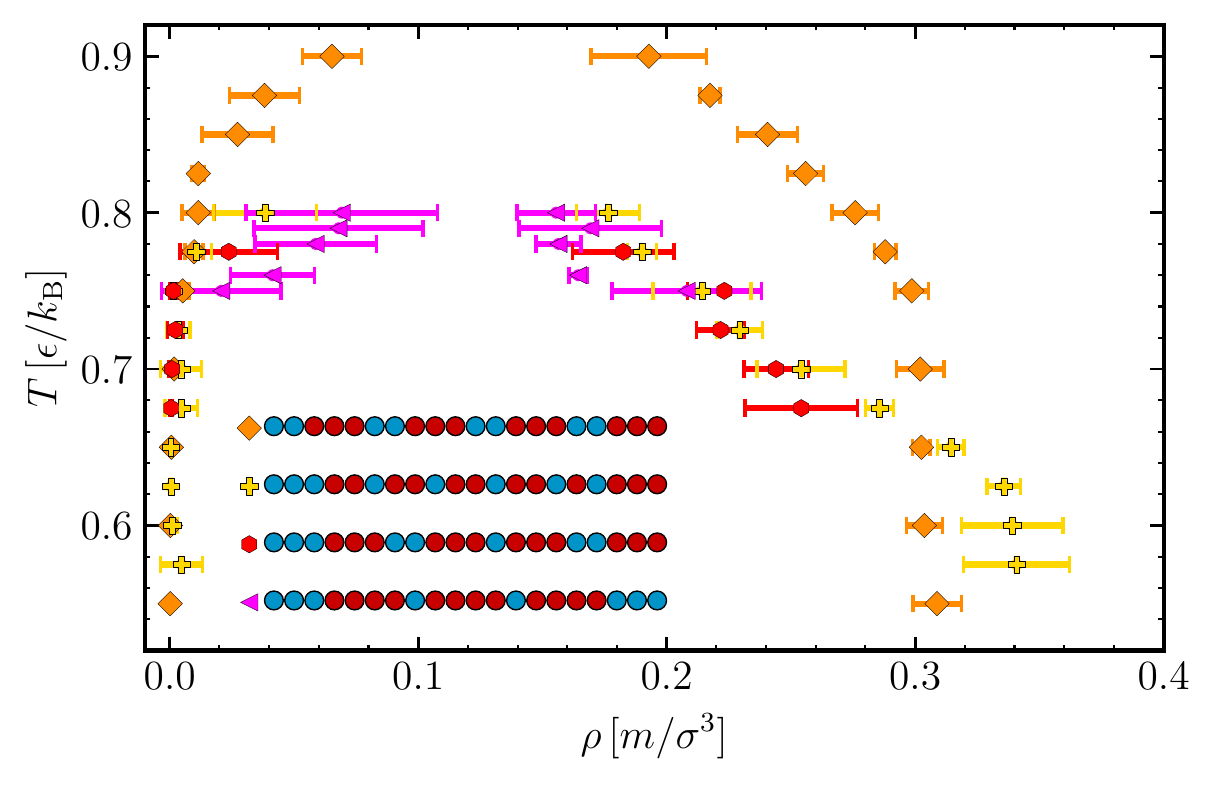}
    \caption{Apparent densities of the dilute and dense phase of sequences with $f_T=0.6$ which do not show conventional phase separation. In the legend, hydrophobic T beads are shown in red (\Tc), hydrophilic H beads are shown in blue (\hc).}
    \label{fig:phase_other}
\end{figure}

In Fig.~\ref{fig:phase_other}, we report the apparent coexistence densities of the fibril-like aggregates that did not exhibit a conventional dilute-dense phase coexistence at any temperature. Each density was obtained by averaging over 5 independent direct coexistence runs with $N=4000$ each, started from different initial configurations.  Because of the presence of large voids, the observed density of the dense phase was much lower than for the liquid phases studied in previous sections. Due to sampling limitations at low temperatures, we only report fibril-like large-scale aggregate densities in Fig.~\ref{fig:phase_other}, where we were able to obtain a consistent density value from independent simulation runs. 

In comparison to conventional liquids, we also observed a much higher heterogeneity within the aggregates, including the formation of large voids and holes as well as microphase separation into H and T rich regions. Consequently, the variance of the local density within the dense phase was much higher for a fibril-like aggregate than for a conventional dense liquid. For a normal liquid, the variance decreases with decreasing temperature, but this trend is inverted for the fibril-like aggregates. For reentrant phase behavior, the variance decreases and then increases with decreasing temperature, as shown in Fig.~\ref{fig:hisvar} in the SM.

\begin{figure}
    \centering
    \includegraphics[width=0.45\textwidth]{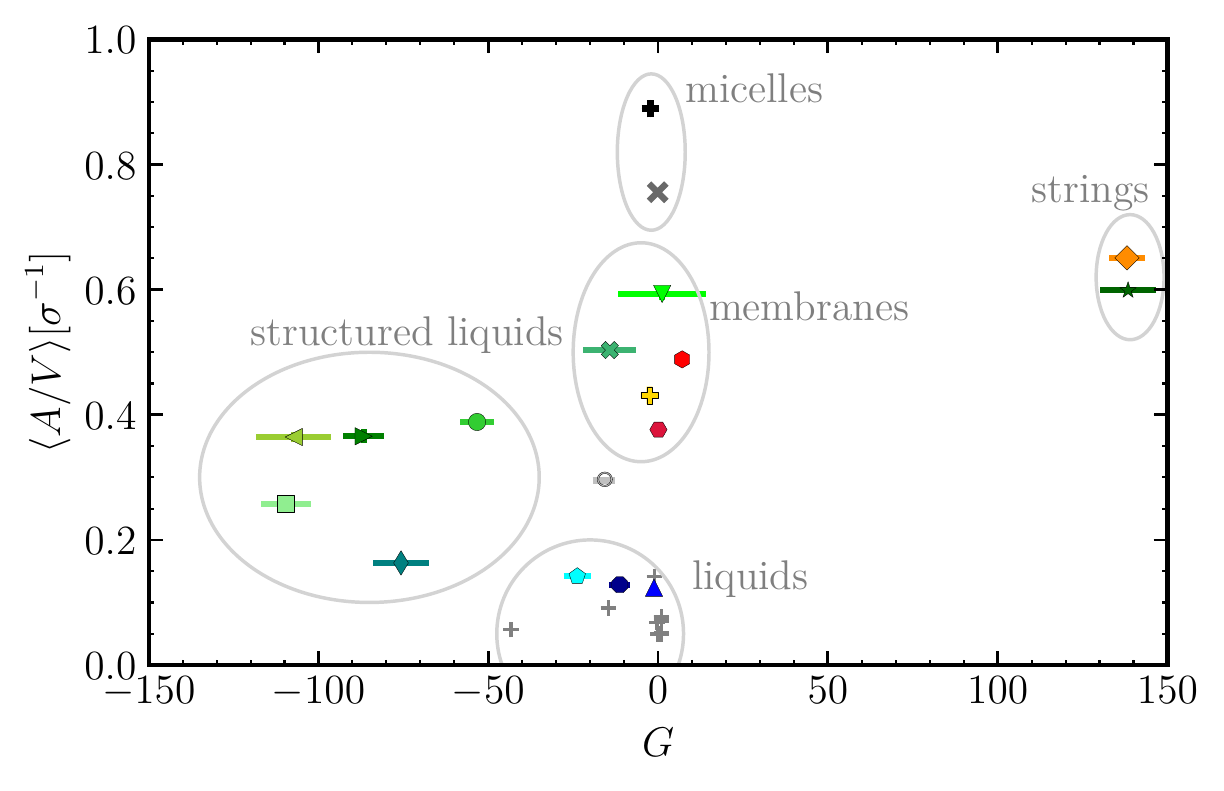} 
    \caption{Genus $G$ and average area-to-volume aspect ratio $\langle A/V\rangle$ of the surfaces determined from the configurations with $\rho_\text{tot}=0.1m/\sigma^3$ and $T=0.55\epsilon/k_\text{B}$. The labeled light grey circles are drawn for a better visualization of the different morphologies as identified by eye from snapshsots of configurations, so boundaries are not rigorous. The grey crosses indicate measured $G$ and $\langle A/V\rangle$ values for sequences with $f_T>0.6$. Corresponding typical snapshots are shown in Fig.~\ref{fig:snapshots_other}.}
    \label{fig:genus}
\end{figure}

We determined the surface to volume ratio of the clusters $\langle A/V \rangle $ with a surface mesh method,~\cite{Stukowski2014} using a probe sphere radius of $R=2\sigma$. The resulting surface is a triangulated mesh enclosing the dense aggregates or liquid phase. As expected, the liquids have the lowest average surface to area ratio, as shown in Fig.~\ref{fig:genus}. The large-scale aggregates are less dense and have holes or voids, leading to a higher surface area. We note that the results were shifted for different temperatures, but the relative ordering of the sequences remained unchanged. 

From the triangulated mesh, we were able to calculate the genus of the surface, $G = 1 -\chi/2$, where $\chi=N_T-N_E+N_V$ is the Euler characteristic.~\cite{Sheth2003} $N_T$ is the number of faces, $N_E$ is the number of edges, and $N_V$ is the number of vertices defining the surface. By determining the genus of the surfaces, we effectively counted the holes (positive $G$) or internal voids (negative $G$). This property, in conjunction with $\langle A/V \rangle$, was able to distinguish between the different large-scale aggregates, as visible in Fig.~\ref{fig:genus}. 

There are no sharp boundaries between the different large-scale aggregate types and some sequences exhibit different types of behavior depending on temperature (e.g. reentrant phase behavior or a string-like to membrane-like transition in $\text{H}_{3}[\text{T}_4\text{H}]_3\text{H}_2$). 
Therefore, the boundaries drawn in Fig.~\ref{fig:genus} serve as guidelines only. We observe that, in contrast to the conventional polymer systems, the sequences investigated here are less regular, so relatively small changes in the sequence have a significant effect on the aggregation behavior.

Analysis of biological condensates using fluorescence microscopy can provide information regarding the localization of IDPs, but microstructure and conformational states of the aggregates are often difficult to access in experimental studies. However, these might be of biological relevance. Microstructure in intracellular condensates has been reported in both P granules,~\cite{Putnam2019} where gel and liquid phases are co-assembled, and stress granules,~\cite{Jain2016} where a network of protein-protein interactions is formed. It has been suggested that microstructures are essential elements for biomolecular condensates,~\cite{Putnam2019} pointing towards potential biological relevance of the large-scale aggregates that form in our simulation.

Pathological protein aggregation also plays an important role in diseases such as ALS and Alzheimer's.~\cite{Chiti2006,Shin2017} 
However, the precise mechanism explaining how point mutations in disordered proteins can give rise to pathological aggregates is not fully understood.~\cite{Chiti2006,Shin2017} We speculate that a systematic classification of the different connections between sequence and aggregate types may lead to advances in both understanding the diseases and, potentially, new drug development. 

\subsection{Order parameter \label{sec:orderparameter}}

As shown in the previous sections, this simple model for IDPs shows diverse phase behavior. We observed conventional dense-dilute phase separation, reentrant phase behavior, and large-scale aggregate formation. To distinguish between all of them and to estimate the critical point location, we would like to have a predictive order parameter. 

We have tested commonly used order parameters: (1) the average length $L_T$ of the hydrophobic segments in the sequence,~\cite{Pandav2012} (2) the normalized mean-square fluctuation of the block hydrophobicity $\Psi^{(s)}$ of a sub-section of length $s$,~\cite{Irback2000,Irback1996} (3) the sequence charge decoration SCD,~\cite{Das2018} (4) the order parameter $\kappa$,~\cite{Das2013} and (5) the corrected probability of finding a T segment after a T segment $P_{TT} - f_T$.~\cite{Flory1955,Shan2007} The SCD and $\kappa$ parameters are used frequently for proteins,~\cite{Das2018} whereas the other order parameters are usually applied to co-polymer systems. Details can be found in section~\ref{sec:orderparameterSI} in the SM.

\begin{figure}
    \centering
    \includegraphics[width=\columnwidth]{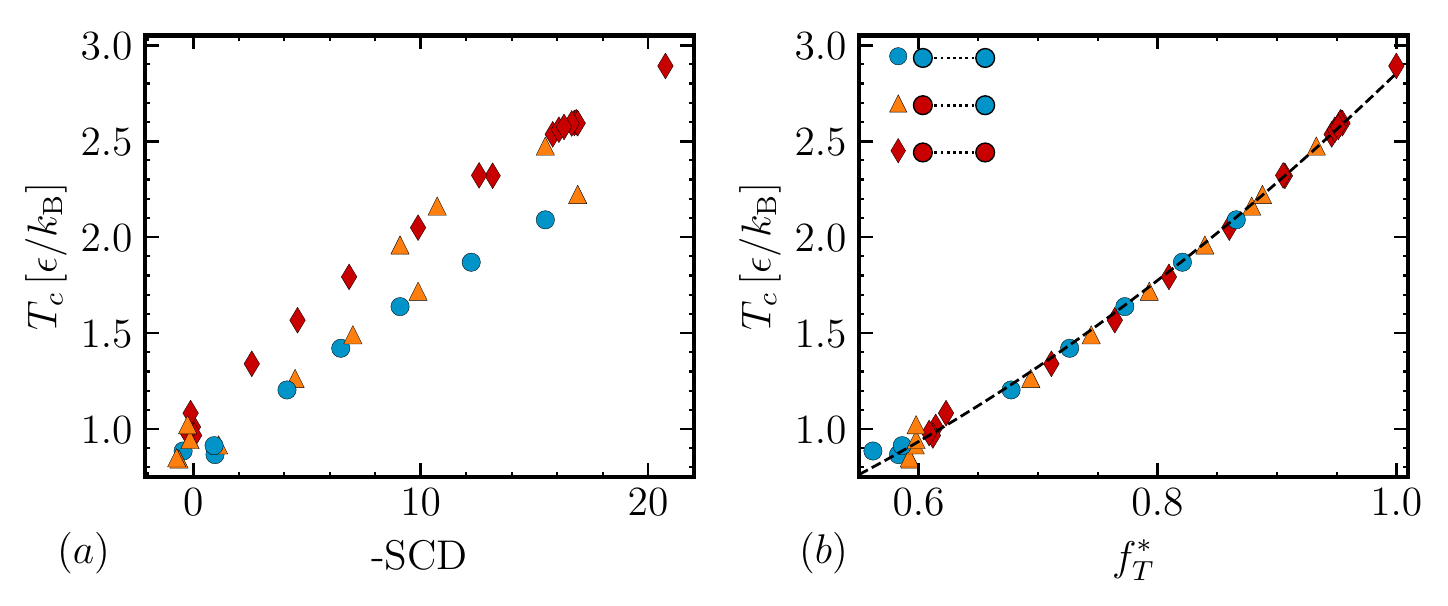} 
    \caption{Critical temperature $T_c$ as a function of (a) sequence charge decoration SCD and (b) reweighted $f^*_T$  for all sequences with a critical point. Red diamonds (\diamond) indicate sequences where both terminal beads are T, blue circles (\circles) mark sequences with two H ends, and orange triangles (\triangle) indicate the sequences with one H and one T end. The dashed line in (b) is a quadratic fit. }
    \label{fig:orderparameter}
\end{figure}

All of the tested order parameters are based only on single chain sequence, which is desirable for predictive capability. Unfortunately, most of them perform poorly (Fig.~\ref{fig:orderparameterSI} in the SM), with SCD showing the best correlation with $T_c$, as can be seen in Fig.~\ref{fig:orderparameter}(a). None of the order parameters in literature take the distance of each bead to the end of the chain into account, and thus cannot predict the strong effect of changing the terminal bead type. For SCD, changing the end bead seems to cause a constant offset, which might allow us to incorporate the terminal bead type effect to this order parameter in the future.

To capture the variation in the effect size of changing the bead type based on the position of the bead in the chain, we defined a new order parameter that acts as an effective reweighted $f^*_T$. The weights of T beads are determined based on the critical temperature of sequences with $f_T=0.95$ (as reported in Fig.~\ref{fig:T19phase}(b) in the SM), while H beads are given zero weight. The full definition can be found in the SM. We have not considered the impact of H beads on entropic effects related to chain configurations, and we  neglect chain length effects that may impact phase separation. 

With this definition, we can achieve a fairly linear correlation between $f^*_T$ and $T_c$ for all investigated sequences which had a critical point, as visible in Fig.~\ref{fig:orderparameter}(b). In fact, it seems to follow the quadratic trend expected for $f_T$ (as in Fig~\ref{fig:crit_scaling}). This definition is specific to the model investigated here and it is not purely based on the sequence alone. Regardless of its limitations, this order parameter illustrates the significance of the beads near the end of the chain for the location of the critical point.

None of the order parameters or single-chain properties tested here, including $f^*_T$, $R_g$, and $T_\Theta$ were able to predict the different observed types of phase behavior (e.g. reentrant) and aggregation (e.g. membrane-like structures) discussed in sections~\ref{sec:reentrant} and~\ref{sec:aggregate}.

From the previous results it is clear that the overall fraction of hydrophobic beads $f_T$, the type of terminal beads and the ``blockiness'' of the sequence are the most important factors to consider for an order parameter. Low blockiness sequences with sufficient fraction of T beads appeared to undergo conventional phase separation. Increasing blockiness seemed to lead to a higher propensity to show reentrant phase behavior, and eventually to the formation of large-scale aggregates. Even higher blockiness resulted in finite sized aggregates like micelles. 




\section{Conclusions ~\label{sec:outlook}}

In this work, we determined the phase behavior of 37 different coarse-grained model IDP sequences. We investigated the influence of both the fraction of hydrophobic beads $f_T$ as well as the distribution of hydrophobic beads along the chain for $f_T=0.6$. We have found phase behavior ranging from conventional dense-dilute phase separation to reentrant phase behavior, as well as large-scale aggregate formation.

Our results show that the types of the beads located at the end of the chain (hydrophobic or hydrophilic) have a systematic effect on the location of the critical point, where hydrophobic terminal beads increase both $T_c$ and $\rho_c$ and hydrophilic terminal beads decrease $T_c$ and $\rho_c$. We then showed that the systematic shift is related to the composition of the interfacial region and speculate that it is intimately connected to the interfacial tension. To our knowledge, this systematic trend has not been reported before.

For many sequences with $f_T=0.6$ we observed reentrant phase behavior, where the density of the liquid phase first increased and then decreased with decreasing temperature. This intriguing complex phase boundary behavior was due to the emerging order in the dense phase upon cooling. For protein systems, this is proposed to be an important mechanism for dissolving membraneless organelles~\cite{Milin2018} and for forming vacuoles.~\cite{Banerjee2017} While the simple one-component model in this work showed reentrant phase behavior, it cannot be easily mapped onto the more complex experimental systems, which are usually driven by salt~\cite{Jordan2014,Zhang2008,Zhang2010} or RNA concentration.~\cite{Banerjee2017,Milin2018} Establishing a qualitative link to biologically relevant systems by introducing some of these effects is outside the scope of this work.  

We also found some sequences which form large-scale aggregates instead of conventional condensed phases. These aggregates differ in their morphology and can be distinguished by their area-to-volume ratio and their number of holes and voids. We observed membrane-like configurations, interconnected fibril-like networks, as well as traditional spherical and worm-like micelles. Seemingly minor changes in the sequence of the model protein led to large changes in the phase behavior. Many intracellular condensates may potentially exhibit rich substructures,~\cite{Jain2016,Putnam2019} which likely reflects one aspect of the complex phase behavior captured in the present simulations.

To our knowledge, there is no predictive order parameter which allows us to distinguish dense-dilute phase separation, reentrant phase behavior, and the different large-scale aggregation types observed. Our order parameter $f^*_T$, single chain properties $R_{ij}$ (or $T_\theta$) and $R_g$, and SCD could predict the critical point location to an extent. However, they were all unable to differentiate between conventional liquid-liquid phase separation and other aggregation behavior. Future efforts will be directed towards understanding which sequence features determine phase behavior and towards developing an order parameter. 

Characterizing the influence of sequence on phase behavior is key to understanding the biological function of phase separation of proteins. The rich structural behavior we observed in this work may be linked to pathological protein aggregation found in diseases like ALS and Alzheimer's.~\cite{Chiti2006,Shin2017} Thus, establishing a link between sequence and aggregation behavior, as done in this study, might lead to insights into the connection between point mutations and pathological aggregation. Further research efforts are needed to better characterize the relation between sequence and phase behavior for more realistic models, which could help understand neurodegenerative diseases and offer insights into drug development.


It would be especially useful to develop a predictive order parameter for biologically relevant protein sequences. With a solid understanding of which sequence features play an important role in phase separation and aggregate formation, engineered condensates could potentially be used for medical applications, as we may be able to predict how specific mutations might change the phase behavior of a given protein.


\section*{Supplementary Material}

See supplementary material for additional information on simulation details, the  scaling of the radius of gyration and the $\Theta$-temperature with the critical point, phase diagrams and interface compositions of the sequences with $f_T=0.95$, and  additional phase diagrams for sequences with $f_T=0.6$. The structure of the liquid and large-scale aggregates, as well as the order parameters used in this work are also discussed. Two tables with all 37 sequences investigated and their critical points are provided. 

\begin{acknowledgments}
We thank Ushnish Rana for valuable comments and discussions.
A.S. was supported by the Princeton Center for Complex Materials (PCCM), a U.S. National Science Foundation Materials Research Science and Engineering Center (Grant No. DMR-1420541).
The simulations were performed using computational resources supported by the Princeton Institute for Computational Science and Engineering (PICSciE) and the Office of Information Technology's High Performance Computing Center and Visualization Laboratory at Princeton University.
\end{acknowledgments}

\bibliography{library}

\counterwithin{figure}{section}
\counterwithin{table}{section}
\renewcommand{\thetable}{S\arabic{table}}  
\renewcommand{\thefigure}{S\arabic{figure}} 
\clearpage
\FloatBarrier
\newpage
\onecolumngrid
\section{Supplementary Material}
\subsection{Smoothing function for the pair potential}

For the LJ pair potential, we applied a smoothing polynomial $s(r)$
\begin{align}
  s(r) =  \begin{cases}
1 \quad& r\leq r_s,\\
 \frac{(r_c^2-r^2)^2(r_c^2+2r^2-3r_c^2)}{(r_c^2-r_s^2)} \quad& r_s < r \leq r_c,\\
0 \quad& r> r_c ,
\end{cases}
\end{align}
which ensures that the pairwise potential and forces transition
smoothly to zero at the truncation radius $r_c$. In this work, we chose $r_c = 3 \sigma$ and began smoothing from $r_s = 2.5 \sigma$.

 \begin{figure}
    \centering
     \includegraphics[width=0.8\textwidth]{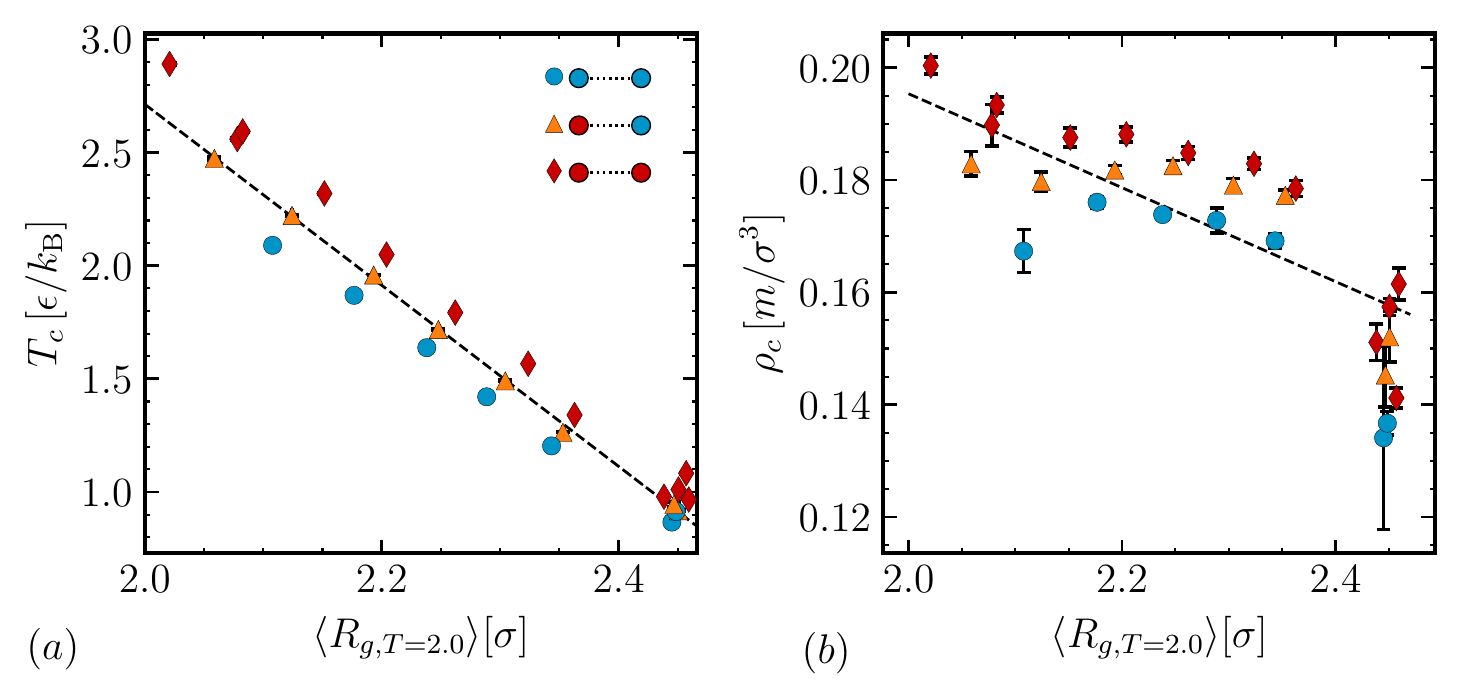}\\
     \includegraphics[width=0.8\textwidth]{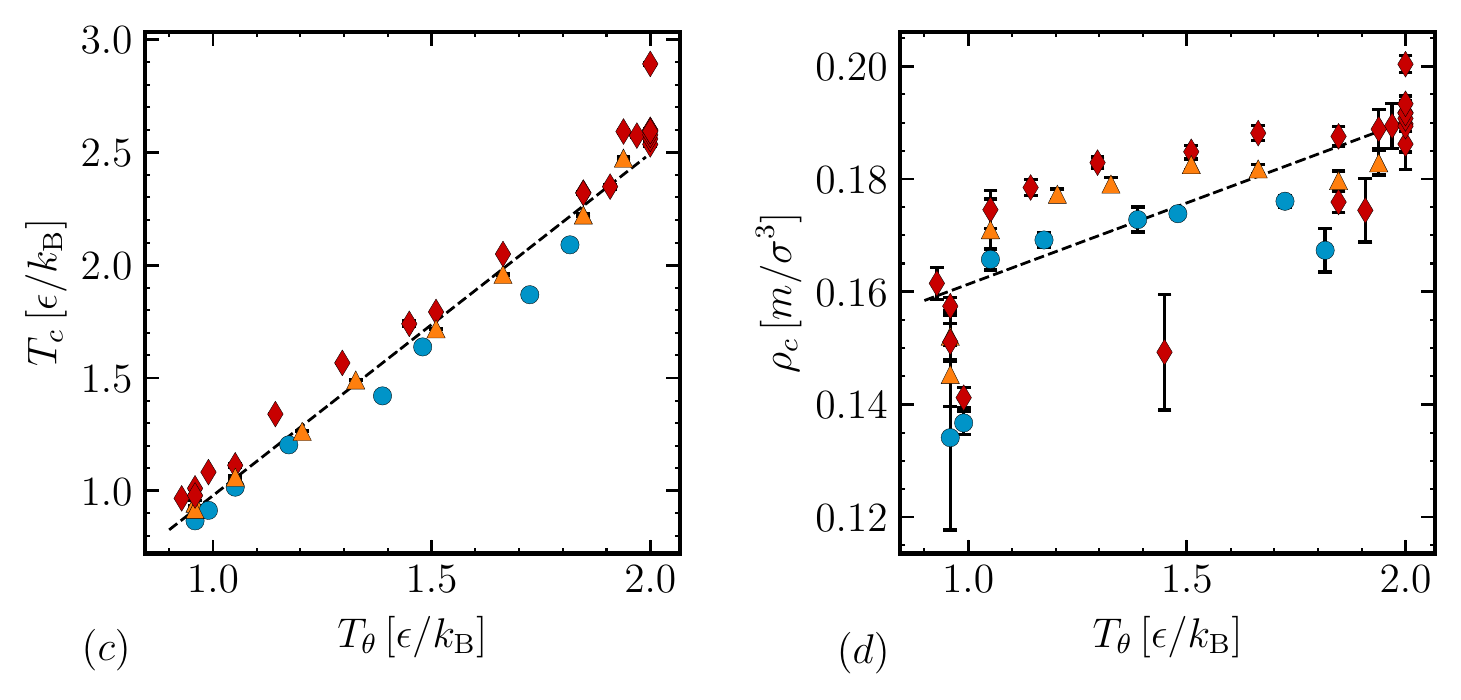}
      \caption{(a) Scaling of the critical temperature $T_c$ with the radius of gyration of a single chain $\langle R_g\rangle$ measured at a fixed temperature of $T=2.0 \epsilon/k_\text{B}$. (b) Same scaling as in (a) for the critical density $\rho_c$. (c) Scaling of $T_c$ with $T_\theta$. (d) Scaling of $\rho_c$ with $T_\theta$. Red diamonds (\diamond) indicate sequences where both terminal beads are T, blue circles (\circles) mark sequences with two H ends, and orange triangles (\triangle) indicate the sequences with mixed terminal beads. The dashed lines are linear fits to the data for sequences with mixed terminal beads.}
    \label{fig:TRgThetaScaling}
\end{figure}

\subsection{Scaling of single-chain properties with $f_T$}

As previously suggested, the single chain radius of gyration at a fixed temperature~\cite{Dignon2018,Lin2017,Lin2017_2} should scale with the critical temperature and density. As shown in Figs.~\ref{fig:TRgThetaScaling}(a) and (b), this holds true for our model as well. While the critical temperature $T_c$ scaled rather well with $R_g$, the correlation with the critical density $\rho_c$ was less  pronounced. 
 
 Since the critical temperature $T_c$ and the $\theta$-temperature $T_\theta$ are also connected~\cite{Panagiotopoulos1998,Dignon2018} and a better scaling relationship with $T_\theta$ for chains of different lengths is expected, we calculated $T_\theta$. We determined the  coil-to-globule transition from the average bead distance $R_{ij}$ between bead $i$ and $j$ along the chain. The temperature where the coil-to-globule occurs is the $\theta$-temperature. For details on this method we refer the reader to \citet{Dignon2018}  The results are shown in Figs.~\ref{fig:TRgThetaScaling}(c) and (d).  
 Similar to $R_g$, $R_{ij}$ showed roughly linear scaling, with better results for $T_c$ than $\rho_c$. 
 
Both $R_g$ and $R_{ij}$ can provide an estimate of the critical point location; however, they fail to capture the systematic influence of the terminal beads found in our simulations. As illustrated in section~\ref{sec:aggregate}, it is also not possible to predict if a certain sequence will phase separate or not by using $R_g$ or $R_{ij}$.

\FloatBarrier
\subsection{Sequences with $f_T=0.95$}

\begin{figure}[h]
    \centering
     \includegraphics[width=\columnwidth]{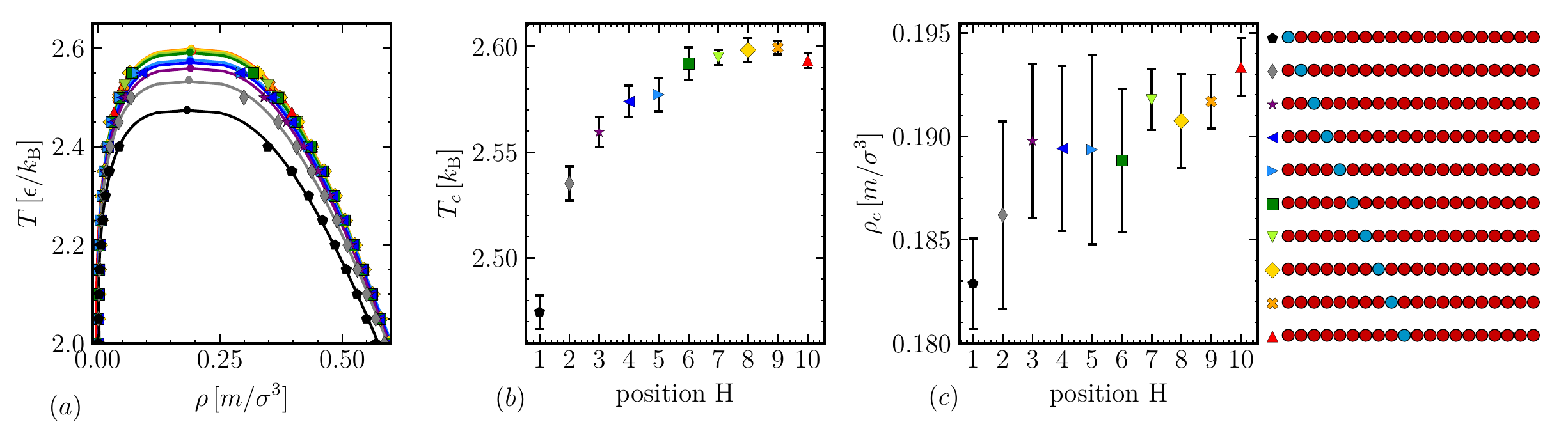}
    \caption{(a) Phase diagrams for sequences with $f_T=0.95$ with varying position of the one hydrophilic bead. The locations of the critical points are indicated by circles; the lines show the fit of Eqs.~(\ref{eq:univ})--(\ref{eq:rect}). (b) Scaling of the critical temperature and (c) of the critical density as a function of the position of the repulsive H (\hc) bead going from the end of the chain (1) to the middle (10).  }
    \label{fig:T19phase}
\end{figure}

Fig.~\ref{fig:T19phase}(a) shows the phase envelopes of all sequences with $f_T=0.95$, and Figs.~\ref{fig:T19phase}(b) and (c) show the scaling of the critical point with the position of the repulsive hydrophilic head, measured as the distance from the end of the chain. 

\begin{figure}
    \centering
    \includegraphics[width=0.9\columnwidth]{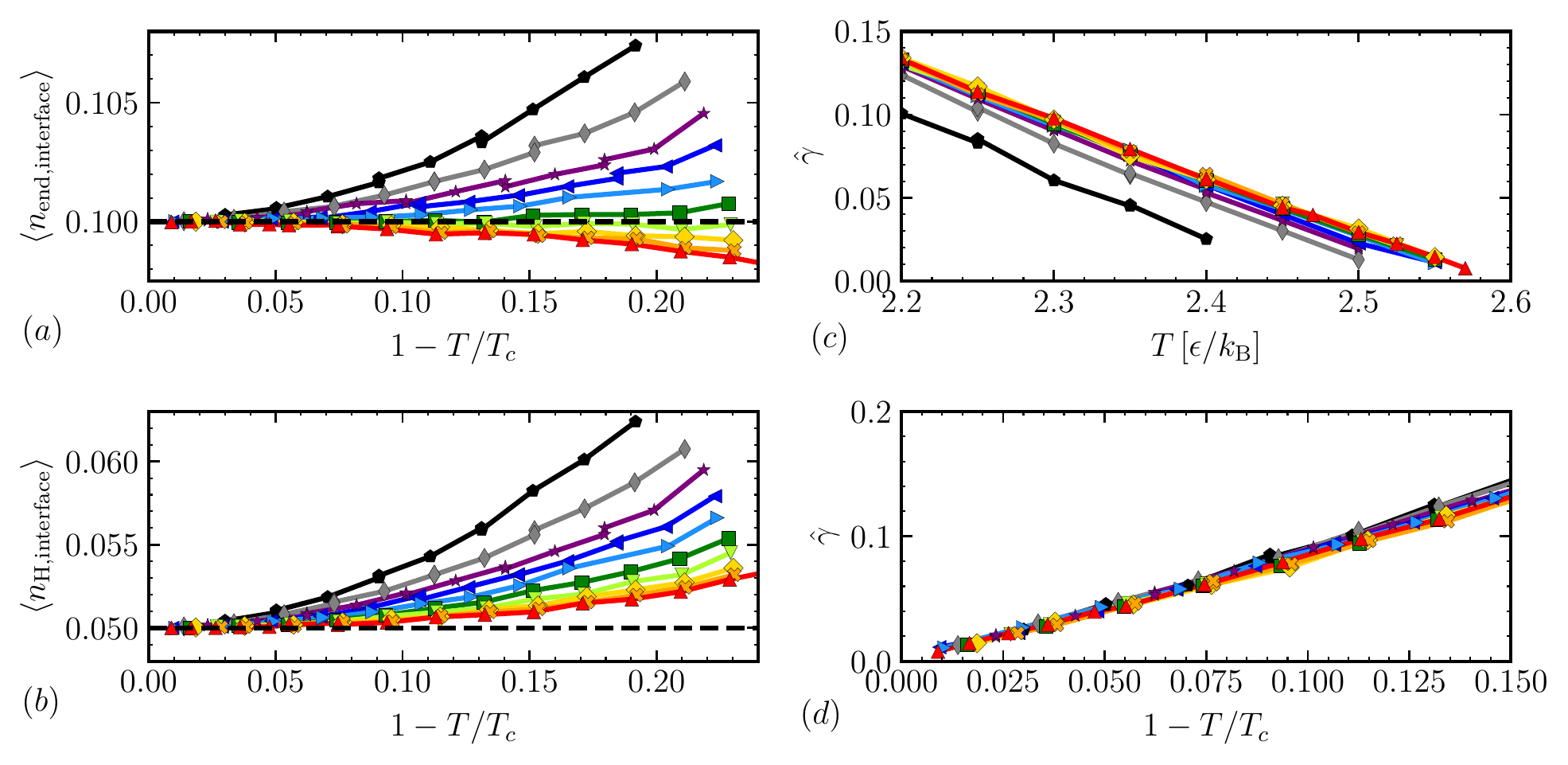}
    \caption{Average number of end beads (a), and average number of H beads (b) in the interfacial region. The dashed lines indicate the average numbers for completely random distributions. $\hat \gamma$ is shown as a function of temperature (c) and as a function of distance to the critical temperature, $1-T/T_c$ (d). The legend is the same as in Fig.~\ref{fig:T19phase}. }
    \label{fig:T19interface}
\end{figure}

As shown in Figs.~\ref{fig:T19interface}(a) and (b), the interfacial composition changes significantly with both temperatures and position of H along the chain. The interface location was found by fitting a $\tanh$ profile to the density histogram. We defined the interfacial region based on the location where the $\tanh$ reached 10\% and 90\% of its bulk density value. For each sequence and at each temperature, the interface of the liquid had a unique composition of H and T beads, as well as a unique relative percentage of end beads. When the hydrophilic H bead is located at the end of the chain, a significant increase in the number of end beads in the interfacial region in comparison to a random distribution was observed below the critical point. When the hydrophobic bead was closer to the center of the chain, we found only a weak enhancement of end beads in the interfacial region. For $T_{19}H$, $T_{18}HT$, and $T_{17}HT_2$, we even observed fewer end beads in the interfacial region than expected from a purely random distribution below the critical point.  

As a rough estimate of the surface tension, we determined $\hat \gamma = \frac{k_\text{B}T}{2\pi w^2} \ln L $ from the interfacial width $w$, as determined by the $\tanh$ fit.~\cite{Senapati2001} Results are shown in Fig.~\ref{fig:T19interface}(c). Here, $L$ is the box size in $x$ and $y$ dimensions. In order to determine the true interfacial tension, we would need the bulk correlation length $l_b$, which is of the order of the molecular size $\sigma$. The value of $\hat \gamma $ varies with temperature and sequence, but it can be rescaled onto one master curve. All sequences collapsed when plotted against the distance to the critical point $1-T/T_c$ of each sequence, as shown in Fig.~\ref{fig:T19interface}(d). This is expected from the scaling law $\gamma = \gamma_0 (1-T/T_c)^\mu$, with $\mu \approx 1.26$ from the Ising Universality class,~\cite{Widom2013} and it illustrates the importance of interface composition. The position of the H bead changes the interfacial composition at a fixed temperature, which, in turn, changes $\hat \gamma$ at that temperature. Because $T_c$ and the interfacial tension are connected, a change in interfacial composition leads to a change in the critical temperature. 

\FloatBarrier
\subsection{Phase diagrams of sequences with $f_T=0.6$}

\begin{figure}[h]
    \centering
    \includegraphics[width=0.9\columnwidth]{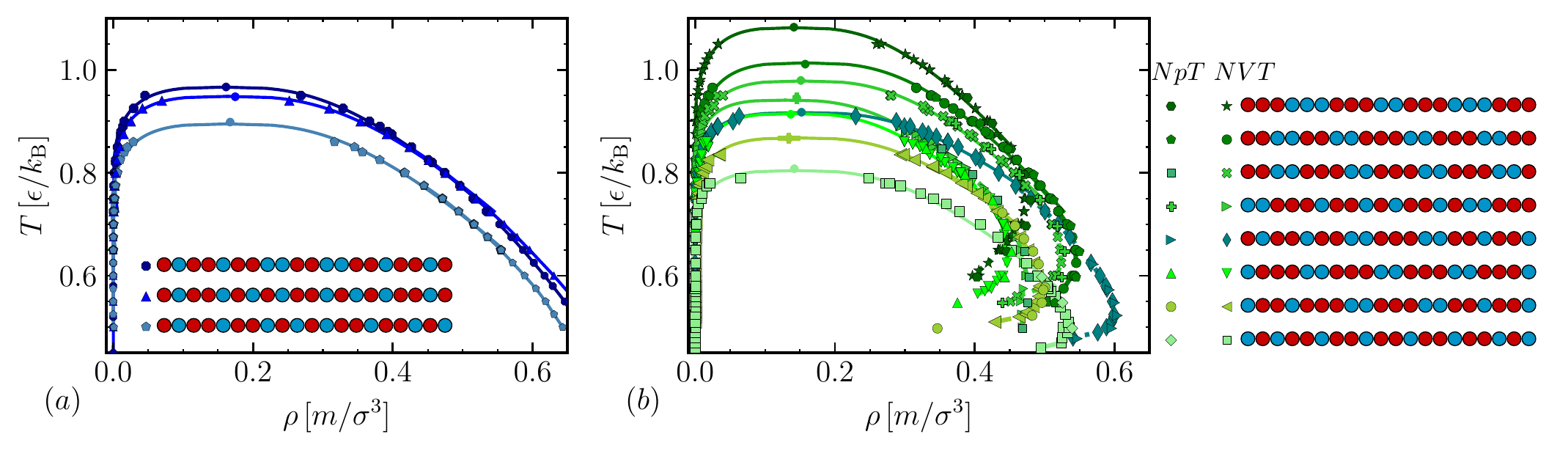}
      \caption{(a) Coexistence densities for sequences with $f_T=0.6$ showing conventional liquid-like phase behavior. (b)  Coexistence densities for all sequences with $f_T=0.6$ showing reentrant phase behavior. The densities were measured in $NVT$ and $NpT$ ensembles as indicated. The locations of the critical points are indicated by circles, and the solid lines show the fit of Eqs.~(\ref{eq:univ})--(\ref{eq:rect}) to the upper part of the $NVT$ phase envelope.}
    \label{fig:T12liquid}
\end{figure}

\begin{figure}
    \centering
    \includegraphics[width=0.95\columnwidth]{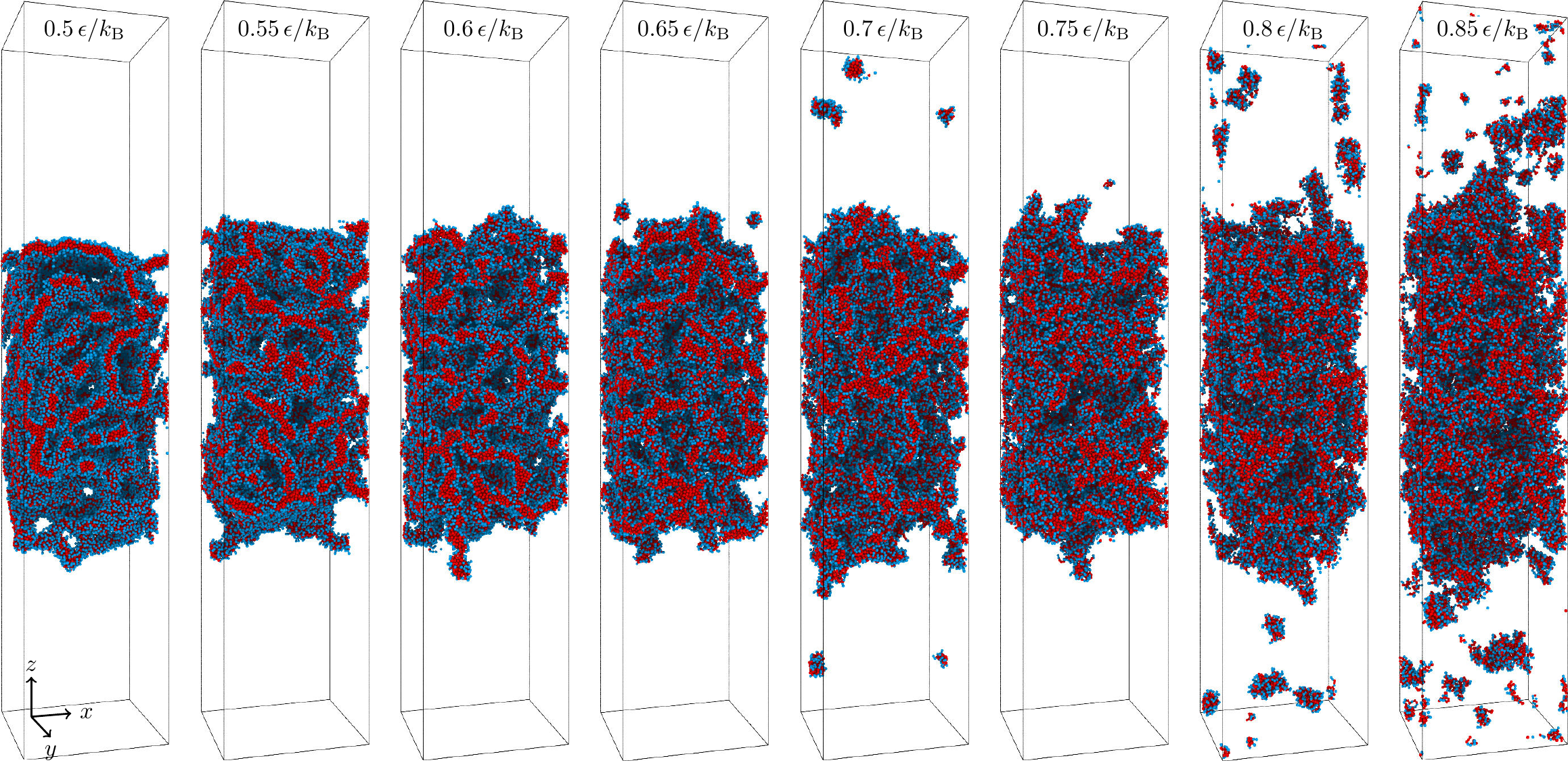}
    \caption{Typical configurations of the regular sequence $[\text{T}_3\text{H}_2]_4$ (\Tc\Tc\Tc\hc\hc\Tc\Tc\Tc\hc\hc\Tc\Tc\Tc\hc\hc\Tc\Tc\Tc\hc\hc) at different temperatures. Snapshots were generated using OVITO~\cite{ovito}.}
    \label{fig:snapshots_strings}
\end{figure}

Fig.~\ref{fig:T12liquid} shows the phase diagrams for all the sequences with $f_T = 0.6$ that showed conventional phase separation or reeentrant behavior. 

\FloatBarrier
\subsection{Structure of the liquid and large-scale aggregates \label{sec:structureliquid}}

\begin{figure}[H]
    \centering
    \includegraphics[width=\columnwidth]{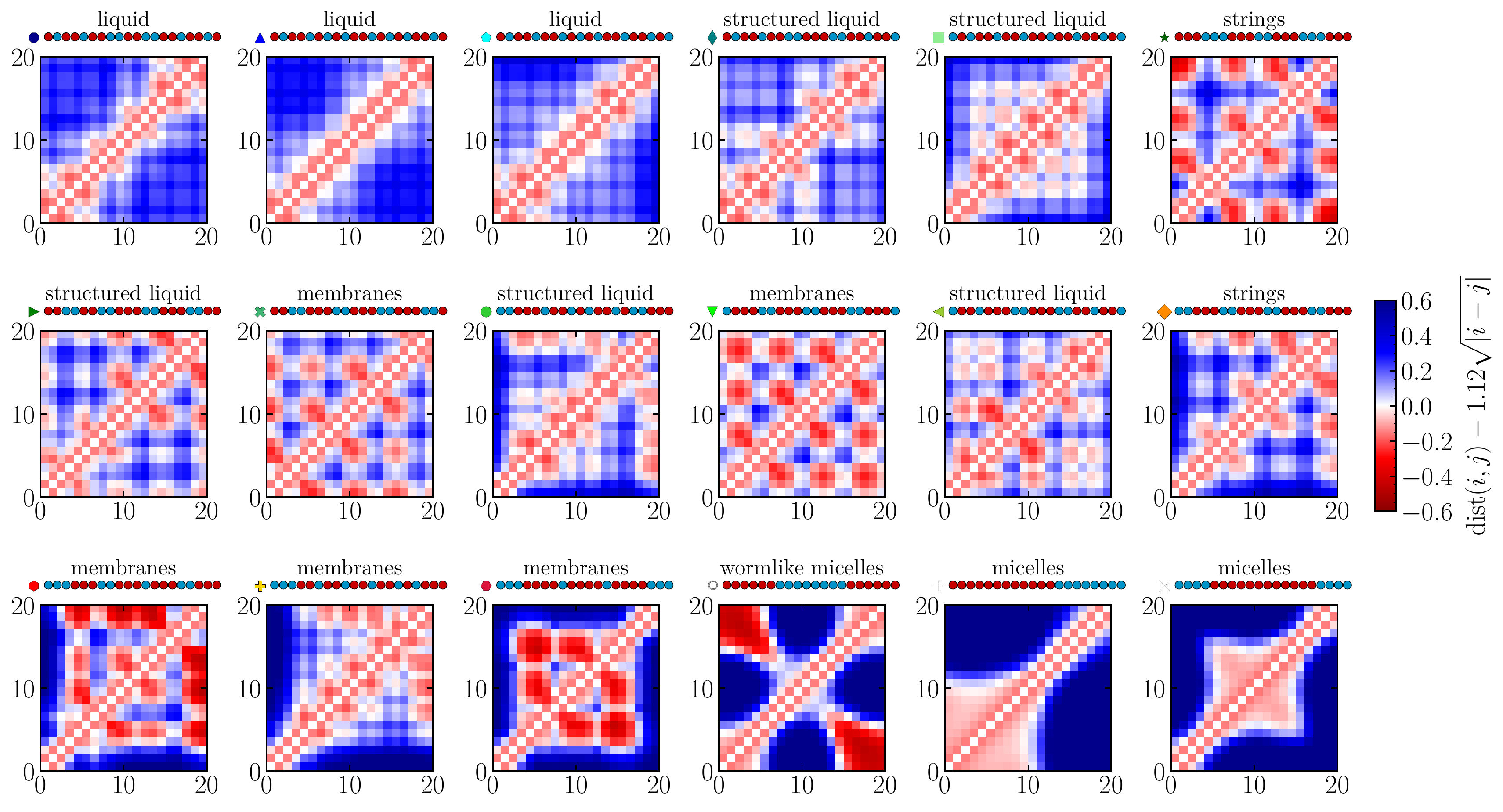}
    \caption{Intra-chain distance maps of bead $i$ to $j$ along a sequence for all sequences with $f_T=0.6$. The density was $\rho_\text{tot}=0.1 m/\sigma^3$ and the temperature was $T = 0.55 \epsilon/k_\text{B}$. The color indicates the deviation of the measured distance from the expected random coil value, $1.12 \sqrt{|i-j|}$. The labels were assigned according to the classification in Fig.~\ref{fig:genus}.}
    \label{fig:dist_maps}
\end{figure}

For ordered proteins, it is common to determine the protein contact map to identify structures like $\alpha$ helices or $\beta$ sheets within the protein. Even though the subject of this study is disordered proteins, these are not completely unstructured, so we determined the intra-chain distances for the different sequences to quantify their degree of ordering. Fig.~\ref{fig:dist_maps} shows the distance $\text{dist}(i,j)$ of monomer $i$ to $j$ in comparison to the expected random coil value for different sequences. Blue indicates a larger than expected distance, red indicates a smaller than expected distance. All sequences except the liquid-like systems show some structure in the distance map at cold temperatures. Aside from the micelle-like systems, it is not possible to unambiguously identify the type of large-scale aggregate from the distance maps. 

\begin{figure}[H]
    \centering
    \includegraphics[width=\columnwidth]{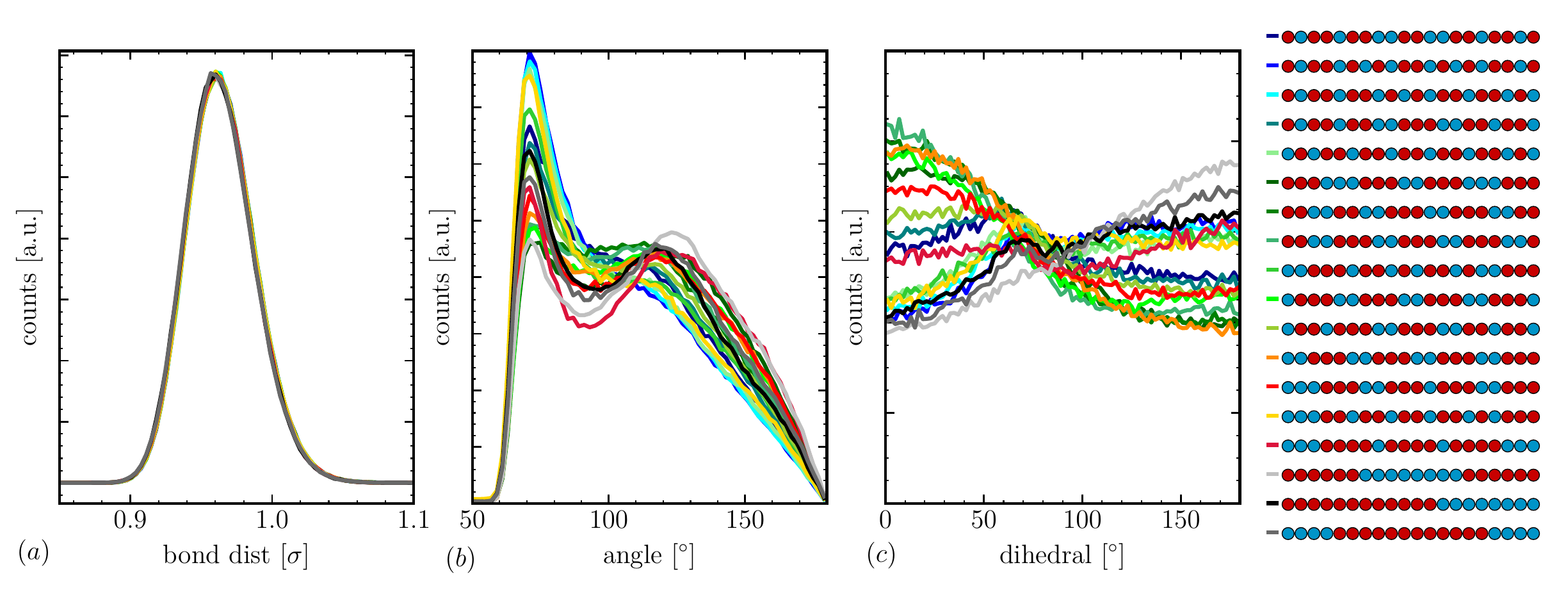}
    \caption{(a) Average bond distance, (b) average bond angle, and (c) dihedral angle distributions of the different sequences with $f_T=0.6$ at $T=0.55 \epsilon/k_\text{B}$ and $\rho_\text{tot}=0.05 m/\sigma^3$.}
    \label{fig:angles}
\end{figure}

\begin{figure}[H]
    \centering
    \includegraphics[width=0.8\columnwidth]{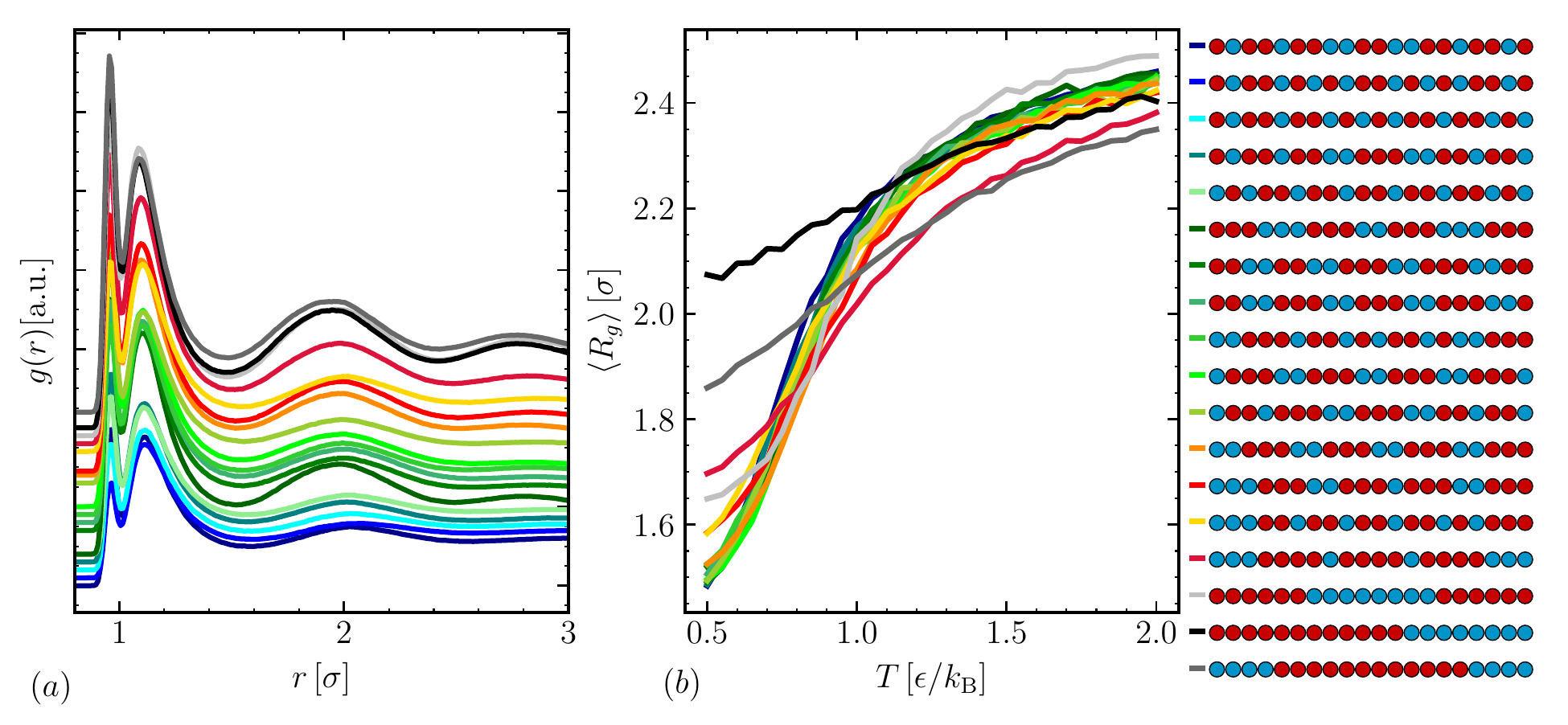}
    \caption{(a) Pair correlation function $g(r)$ for all the sequences with $f_T=0.6$ at $T=0.55 \epsilon/k_\text{B}$ and $\rho_\text{tot}=0.05 m/\sigma^3$. The pair correlation functions are shifted horizontally for clarity. (b) Average radius of gyration $\langle R_g\rangle $ of an isolated chain as function of temperature.}
    \label{fig:rgtemp}
\end{figure}

\begin{figure}[H]
    \centering
    \includegraphics[width=0.65\textwidth]{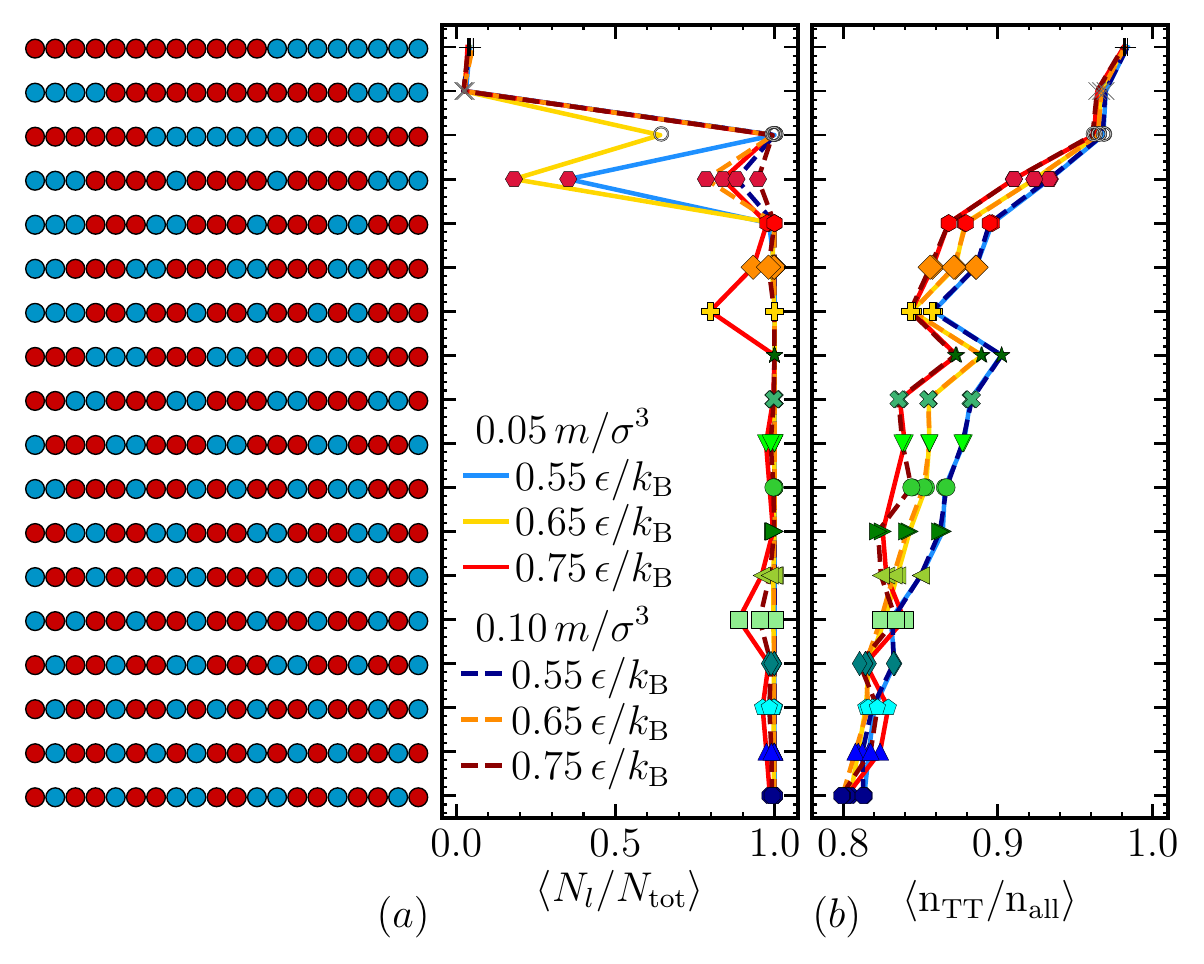} 
    \caption{(a) Average fraction of chains in the largest cluster $\langle N_l/N_\text{tot}\rangle $ and (b) average normalized number of T neighbors of a T bead $\langle\text{n}_\text{TT}/\text{n}_\text{tot}\rangle$. The different lines correspond to three temperatures for $\rho_\text{tot}=0.05\,m/\sigma^3$ (solid) and $\rho_\text{tot}=0.1\,m/\sigma^3$ (dashed). The $y$-axis represents all sequences with $f_T=0.6$, with T beads in red (\Tc) and H beads in blue (\hc).}
    \label{fig:cluster_prop}
\end{figure}

Other short-ranged chain properties like average bond distances, angles and dihedral angles did not show distinctive features for the different large-scale aggregates, as visible in Figs.~\ref{fig:angles}(a)--(c). Inter-chain structural properties like the pair correlation function, shown in Fig.~\ref{fig:rgtemp}(a), also did not reveal a clear way to identify the aggregates, and neither did the radius of gyration, which we determined for a single chain as a function of temperature, shown in Fig.~\ref{fig:rgtemp}(b). 
We measured the properties of the clusters of the system for all sequences forming aggregates. Fig.~\ref{fig:cluster_prop}(a) shows results for the average largest cluster size in the system. In all cases except for the systems that form isolated spherical micelles ($\text{T}_{12}\text{H}_8$ 
and $\text{H}_4\text{T}_{12}\text{H}_4$)  
and a membrane-forming system ($\text{H}_{3}[\text{T}_4\text{H}]_3\text{H}_2$) 
most chains condensed into one single large cluster at both densities and all investigated temperatures, as indicated by $\langle N_l / N_\text{tot}\rangle \approx 1$ in Fig.~\ref{fig:cluster_prop}(a).

Fig.~\ref{fig:snapshots_reentrant} and \ref{fig:snapshots_other} in the main text showed that the different morphologies had a different degree of microphase separation between the hydrophobic and hydrophilic beads. To quantify this effect, we calculated the average number of T neighbors within a distance of $2\sigma$ of T beads $\text{n}_\text{TT}$. This quantity, normalized by the total number of neighbors $\text{n}_\text{tot}$, is shown in Fig.~\ref{fig:cluster_prop}(b). Note that while the absolute value of $\langle \text{n}_\text{TT}/\text{n}_\text{all} \rangle$ shifted for different temperatures, the ordering of the different sequences remained the same. While we observed a rough trend where liquids showed the lowest and micelles showed the highest values of microphase separation, we cannot clearly distinguish different types of large-scale aggregates.

\begin{figure}[H]
    \centering
    \includegraphics[width=0.85\columnwidth]{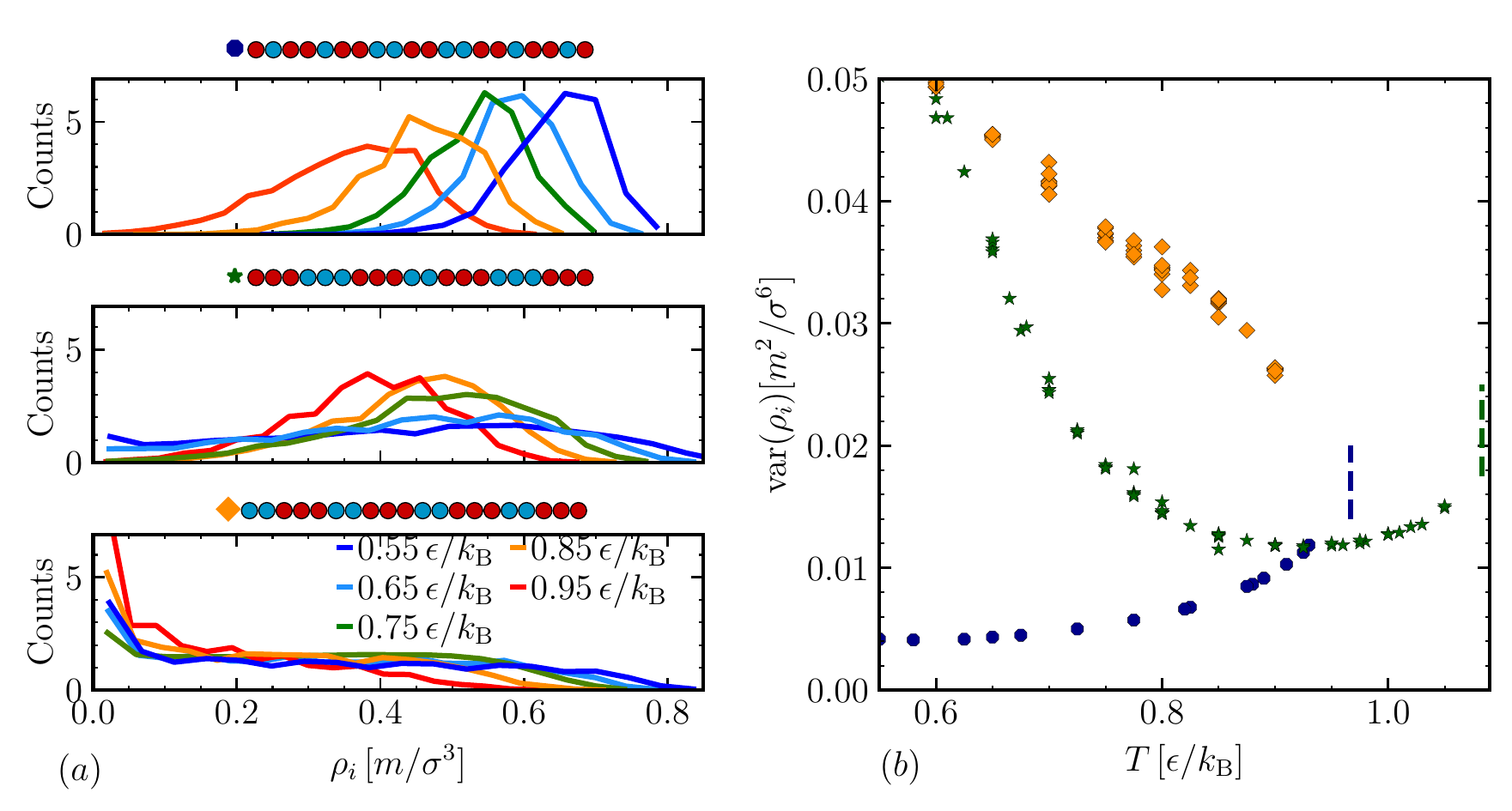}
    \caption{(a) Histograms of local densities $\rho_i$ in $5\sigma^3$ sub-regions within liquid phase at five different temperatures. The three panels correspond to sequences with: a conventional condensed phase (top), a reentrant phase (middle), and string-like aggregates (bottom). (b) Variance of the local density histograms as a function of temperature for the same three sequences. Dashed lines indicate the location of the critical temperatures of the sequences with conventional and reentrant phase behavior.}
    \label{fig:hisvar}
\end{figure}

An interesting difference between the conventional condensed phases and the large-scale aggregates is the variance of local densities. We divided the condensed phase into smaller subsections of size $5\sigma \times 5\sigma \times 5\sigma$ and determined the density $\rho_i$ in each. The resulting histograms for different temperatures are shown in Fig.~\ref{fig:hisvar}(a) for three example sequences. These results hold true for all studied sequences. For a conventional condensed phase (top panel), the distribution of densities in the dense phase shifts to higher densities and becomes narrower as temperature decreases, as expected. This can also bee seen in Fig.~\ref{fig:hisvar}(b), where we plot the variance of the measured densities $\rho_i$ as a function of temperature. For reentrant phase behavior (middle panel), the distributions first shift like in the conventional case, but then flatten out. Some sub-regions contained only as small amount of chains or no chains at all, with $\rho_i\approx 0$, while others were dense. This is a reflection of the formation of fibril-like structures, where some regions in space are very dense and others have large voids. Consequently, the variance first decreases and then increases with lowering the temperature.  For systems that only formed large-scale aggregates (bottom panel), all histograms look similar and do not have a distinguishable peak at a finite density. The variance is high and increases with decreasing temperature, the opposite of a conventional liquid. This behavior can be explained by the increasing degree of microphase separation and the formation of more and more large-scale voids and fibril-like aggregates, making the system increasingly heterogeneous at lower temperatures.

\FloatBarrier
\subsection{Order parameters \label{sec:orderparameterSI}}

\begin{figure}[H]
    \centering
    \includegraphics[width=0.9\columnwidth]{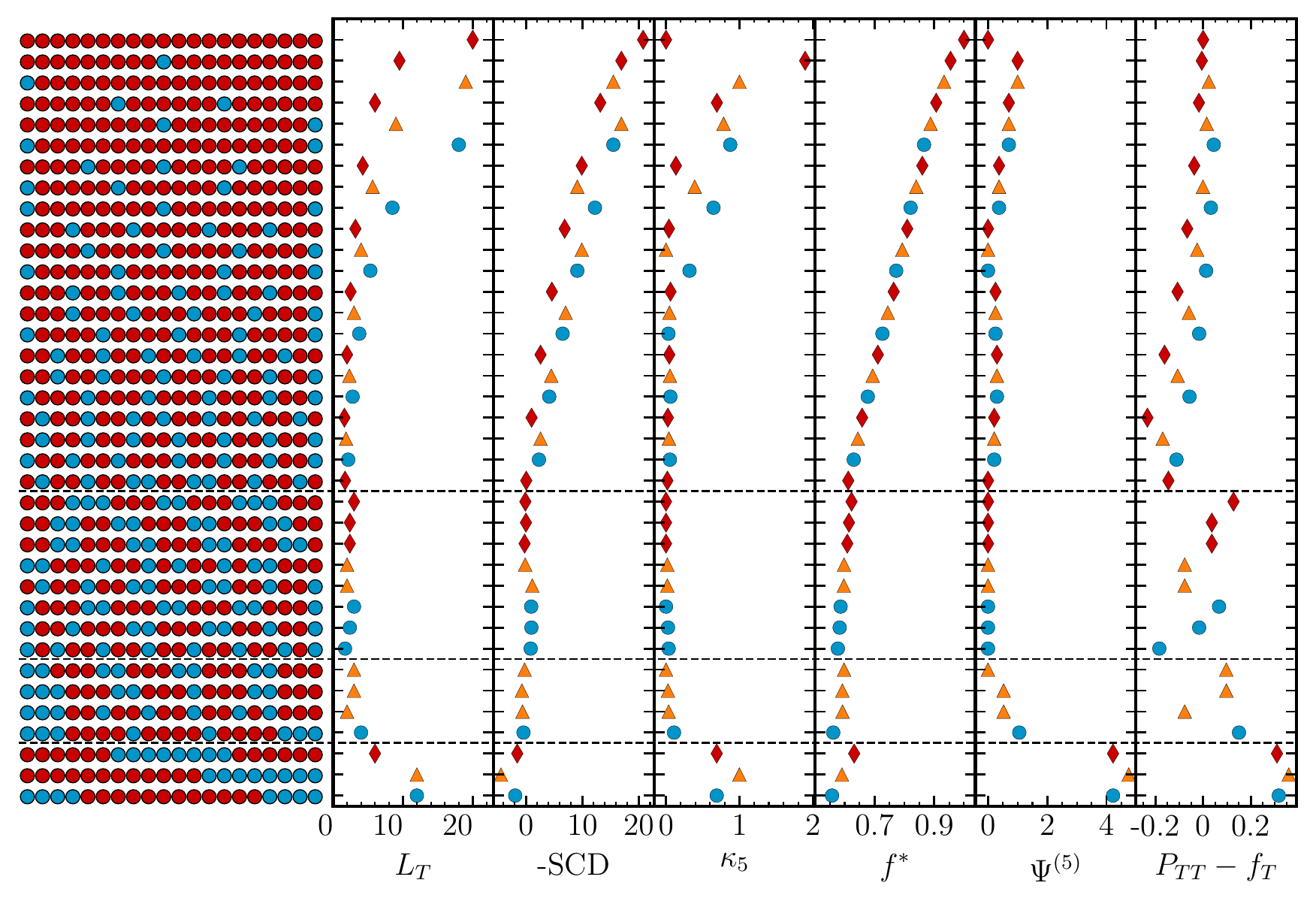} 
    \caption{Red diamonds (\diamond) indicate sequences where both terminal beads are T, blue circles (\circles) mark sequences with two H ends, and orange triangles (\triangle) indicate the sequences with mixed terminal beads. The dashed vertical lines separate sequences which form finite aggregates (bottom), then infinite-range aggregates, then sequences that show reentrant phase behavior, and finally sequences that show conventional dense-dilute phase separation.}
    \label{fig:orderparameterSI}
\end{figure}

In this work, we have tested multiple order parameters. All results are presented in Fig.~\ref{fig:orderparameterSI}. We have chosen to sort the sequences according to their phase behavior (separated by horizontal dashed lines) and then within each group according to their critical temperature, if known. The $y$-axis in Fig.~\ref{fig:orderparameterSI} indicates each sequence in order.  

The first and simplest choice of order parameter is the the average length $L_T$ of the hydrophobic segments in the sequence,~\cite{Pandav2012} shown in Fig.~\ref{fig:orderparameterSI}. This order parameter roughly sorts the sequences, but does not capture the systematic influence of the terminal beads. We also expect that  $L_T$ will not perform well for comparing multiple chain lengths. 

We studied the normalized mean-square fluctuation $\Psi^{(s)}$ of a sub-section of length $s$.~\cite{Irback2000,Irback1996} We varied the length of the sub-sections/blocks $s$ and achieved best results with $s=5$. 
With a few exceptions, $\Psi^{(5)}$ sorts the sequences according to their critical temperature, but cannot be used to distinguish different phase and aggregation behavior, as it only seems to separate out micelle-forming systems.

The order parameter $\kappa$,~\cite{Das2013} commonly used for proteins, does not perform well in this system. We picked a residue blob size of $5$. While $\kappa$ can distinguish different trends within a set of sequences with the same $f_T$, it does not perform as well across different values of $f_T$. We suspect that the reason for this is that $\kappa$ is normalized to its maximal value, which depends on $f_T$. To unify our choice for all values of $f_T$, we normalized against the $\kappa$ value of a corresponding block of composition $f_T$. It is possible that a different choice of normalization might improve results. 

We also determined the (corrected) probability of finding a T segment after a T segment, $P_{TT} - f_T$, where values of $P_{TT} - f_T\approx 0$ correspond to random sequences, $P_{TT} - f_T<0$ to alternating, and   $P_{TT} - f_T>0$ to blocky sequences.~\cite{Flory1955,Shan2007} We observed both negative and positive values within each type of phase behavior except for the sequences that formed finite-sized aggregates, which were all classified as blocky. 

The sequence charge decoration (SCD) is the commonly used order parameter that performed best out of the ones we studied. However, it did not reproduce the systematic influence of the terminal beads. SCD, like the $\kappa$ parameter, is frequently used for proteins,~\cite{Das2018} as opposed to the other studied parameters, which are usually applied to co-polymer systems. 

Since the order parameters from the literature generally performed poorly in our system, we defined an effective reweighted $f^*_T$ as 
\begin{align}
    f^*_T = \frac{\sum^M_{i=1} t_i}{1 - \sum^M_{i=1} \frac{T^{0.95}_c(i)}{T^{1.00}_c}}, \quad
t_i=\begin{cases}
1 - \frac{T^{0.95}_c(i)}{T^{1.00}_c} &\text{ for $i$ = T}\\
0 &\text{ for $i$ = H}
\end{cases}
\end{align}
where $T^{1.00}_c$ is the critical temperature of the pure homopolymer and $T^{0.95}_c(i)$ is the critical temperature of the chain with only one H at position $i$. We have given H beads zero weight because they do not have an energetic contribution to the phase separation, and T beads a weight that depends on their position in the chain. 

By defining $f^*_T$ as described above, we were able to account for the fact that hydrophobic beads closer to the ends of the chain appear to be more important in promoting phase separation. We achieved a fairly linear correlation between $f^*_T$ and $T_c$ for all investigated sequences which had a critical point, as visible in Fig.~\ref{fig:orderparameter}. This definition is specific to the model investigated here and it is not purely based on the sequence alone. Regardless of its limitations, this order parameter illustrates the significance of the terminal beads for the location of the critical point, because we were able to account for their effect with reweighting.

\begin{table}
	\centering
		\caption{Sequence names, architectures, critical point location $T_c$ and $\rho_c$, and fraction of attractive beads $f_T$. Subscript indicates error on the last digit.}
\begin{tabular}{ccccc} 	
		sequence & architecture    &      $T_c$      &  	$\rho_c$  &   	$f_T$\\
		\toprule
$\text{T}_{20}$ & \Tc\Tc\Tc\Tc\Tc\Tc\Tc\Tc\Tc\Tc\Tc\Tc\Tc\Tc\Tc\Tc\Tc\Tc\Tc\Tc & $2.892_5$ & $0.200_2$ & 1.0 \\
\midrule
$\text{H}\text{T}_{19}$ & \hc\Tc\Tc\Tc\Tc\Tc\Tc\Tc\Tc\Tc\Tc\Tc\Tc\Tc\Tc\Tc\Tc\Tc\Tc\Tc & $2.475_8$ & $0.183_2$ & 0.95 \\
$\text{T}\text{H}\text{T}_{18}$ & \Tc\hc\Tc\Tc\Tc\Tc\Tc\Tc\Tc\Tc\Tc\Tc\Tc\Tc\Tc\Tc\Tc\Tc\Tc\Tc & $2.535_9$ & $0.186_5$ 
& 0.95 \\
$\text{T}_{2}\text{H}\text{T}_{17}$ & \Tc\Tc\hc\Tc\Tc\Tc\Tc\Tc\Tc\Tc\Tc\Tc\Tc\Tc\Tc\Tc\Tc\Tc\Tc\Tc & $2.561_7$ & 
$0.189_4$ & 0.95 \\
$\text{T}_{3}\text{H}\text{T}_{16}$ & \Tc\Tc\Tc\hc\Tc\Tc\Tc\Tc\Tc\Tc\Tc\Tc\Tc\Tc\Tc\Tc\Tc\Tc\Tc\Tc & $2.575_8$ & 
$0.189_4$ & 0.95 \\
$\text{T}_{4}\text{H}\text{T}_{15}$ & \Tc\Tc\Tc\Tc\hc\Tc\Tc\Tc\Tc\Tc\Tc\Tc\Tc\Tc\Tc\Tc\Tc\Tc\Tc\Tc & $2.577_9$ & 
$0.189_5$ & 0.95 \\
$\text{T}_{5}\text{H}\text{T}_{14}$ & \Tc\Tc\Tc\Tc\Tc\hc\Tc\Tc\Tc\Tc\Tc\Tc\Tc\Tc\Tc\Tc\Tc\Tc\Tc\Tc & $2.593_7$ & 
$0.189_3$ & 0.95 \\
$\text{T}_{6}\text{H}\text{T}_{13}$ & \Tc\Tc\Tc\Tc\Tc\Tc\hc\Tc\Tc\Tc\Tc\Tc\Tc\Tc\Tc\Tc\Tc\Tc\Tc\Tc & $2.595_4$ & 
$0.192_2$ & 0.95 \\
$\text{T}_{7}\text{H}\text{T}_{12}$ & \Tc\Tc\Tc\Tc\Tc\Tc\Tc\hc\Tc\Tc\Tc\Tc\Tc\Tc\Tc\Tc\Tc\Tc\Tc\Tc & $2.598_5$ & 
$0.191_2$ & 0.95 \\
$\text{T}_{8}\text{H}\text{T}_{11}$ & \Tc\Tc\Tc\Tc\Tc\Tc\Tc\Tc\hc\Tc\Tc\Tc\Tc\Tc\Tc\Tc\Tc\Tc\Tc\Tc & $2.599_3$ & 
$0.192_1$ & 0.95 \\
$\text{T}_{9}\text{H}\text{T}_{10}$ & \Tc\Tc\Tc\Tc\Tc\Tc\Tc\Tc\Tc\hc\Tc\Tc\Tc\Tc\Tc\Tc\Tc\Tc\Tc\Tc & $2.599_2$ & 
$0.1939_9$ & 0.95 \\
\midrule
$\text{H}\text{T}_{18}\text{H}$ & \hc\Tc\Tc\Tc\Tc\Tc\Tc\Tc\Tc\Tc\Tc\Tc\Tc\Tc\Tc\Tc\Tc\Tc\Tc\hc & $2.090_7$ & $0.167_4$ 
& 0.9 \\
$[\text{T}_{6}\text{H}]_2\text{T}_{6}$ & \Tc\Tc\Tc\Tc\Tc\Tc\hc\Tc\Tc\Tc\Tc\Tc\Tc\hc\Tc\Tc\Tc\Tc\Tc\Tc & $2.319_4$ & 
$0.188_2$ & 0.9 \\
$[\text{T}_{9}\text{H}]_2$ & \Tc\Tc\Tc\Tc\Tc\Tc\Tc\Tc\Tc\hc\Tc\Tc\Tc\Tc\Tc\Tc\Tc\Tc\Tc\hc & $2.223_8$ & $0.175_3$ & 0.9 
\\
\midrule
$[\text{T}_{6}\text{H}]_2\text{T}_{5}\text{H}$ & \hc\Tc\Tc\Tc\Tc\Tc\hc\Tc\Tc\Tc\Tc\Tc\Tc\hc\Tc\Tc\Tc\Tc\Tc\Tc & 
$1.959_2$ & $0.1818_8$ & 0.85 \\
$\text{H}\text{T}_{8}\text{H}\text{T}_{9} \text{H}$ & \hc\Tc\Tc\Tc\Tc\Tc\Tc\Tc\Tc\hc\Tc\Tc\Tc\Tc\Tc\Tc\Tc\Tc\Tc\hc & 
$1.869_2$ & $0.1760_1$ & 0.85 \\
$[\text{T}_{4}\text{H}]_3\text{T}_5$ & \Tc\Tc\Tc\Tc\hc\Tc\Tc\Tc\Tc\hc\Tc\Tc\Tc\Tc\hc\Tc\Tc\Tc\Tc\Tc & $2.049_2$ & 
$0.188_1$ & 0.85 \\
\midrule
$\text{H}\text{T}_{5}\text{H}\text{T}_6\text{H}\text{T}_{5}\text{H}$ & 
\hc\Tc\Tc\Tc\Tc\Tc\hc\Tc\Tc\Tc\Tc\Tc\Tc\hc\Tc\Tc\Tc\Tc\Tc\hc & $1.637_2$ & $0.1837_7$ & 0.8 \\
$[\text{T}_{3}\text{H}]_2\text{T}[\text{T}_{3}\text{H}]_2\text{T}_3$ & 
\Tc\Tc\Tc\hc\Tc\Tc\Tc\hc\Tc\Tc\Tc\Tc\hc\Tc\Tc\Tc\hc\Tc\Tc\Tc & $1.792_3$ & $0.187_1$ & 0.8 \\
$[\text{T}_{4}\text{H}]_4$ & \Tc\Tc\Tc\Tc\hc\Tc\Tc\Tc\Tc\hc\Tc\Tc\Tc\Tc\hc\Tc\Tc\Tc\Tc\hc & $1.716_4$ & $0.186_1$ & 0.8 
\\
\midrule
$[\text{H}\text{T}_{4}]_2\text{H}\text{T}_{3}\text{H}\text{T}_{4}\text{H}$ & 
\hc\Tc\Tc\Tc\Tc\hc\Tc\Tc\Tc\Tc\hc\Tc\Tc\Tc\hc\Tc\Tc\Tc\Tc\hc & $1.425_2$ & $0.1818_7$ & 0.75 \\
$[\text{T}_{3}\text{H}\text{T}_{2}\text{H}]_2\text{T}_{2}\text{H}\text{T}_{3}$ & 
\Tc\Tc\Tc\hc\Tc\Tc\hc\Tc\Tc\Tc\hc\Tc\Tc\hc\Tc\Tc\hc\Tc\Tc\Tc & $1.567_2$ & $0.1830_9$ & 0.75 \\
$[\text{T}_{3}\text{H}]_5$ & \Tc\Tc\Tc\hc\Tc\Tc\Tc\hc\Tc\Tc\Tc\hc\Tc\Tc\Tc\hc\Tc\Tc\Tc\hc & $1.491_3$ & $0.179_1$ & 
0.75 \\
\midrule
$[\text{H}\text{T}_{3}]_4\text{H}\text{T}_2\text{H}$ & \hc\Tc\Tc\Tc\hc\Tc\Tc\Tc\hc\Tc\Tc\Tc\hc\Tc\Tc\Tc\hc\Tc\Tc\hc & 
$1.208_1$ & $0.1777_7$ & 0.7 \\
$[\text{T}_{2}\text{H}]_6\text{T}_2$ & \Tc\Tc\hc\Tc\Tc\hc\Tc\Tc\hc\Tc\Tc\hc\Tc\Tc\hc\Tc\Tc\hc\Tc\Tc & $1.340_2$ & 
$0.178_1$ & 0.7 \\
$[\text{T}_{2}\text{H}]_2[\text{T}_{3}\text{H}]_2[\text{T}_{2}\text{H}]_2$ & 
\Tc\Tc\hc\Tc\Tc\hc\Tc\Tc\Tc\hc\Tc\Tc\Tc\hc\Tc\Tc\hc\Tc\Tc\hc & $1.264_2$ & $0.1773_9$ & 0.7 \\
\midrule
$[\text{H}\text{T}_2]_3\text{T}[\text{H}\text{T}_2]_3\text{H}$ & 
\hc\Tc\Tc\hc\Tc\Tc\hc\Tc\Tc\Tc\hc\Tc\Tc\hc\Tc\Tc\hc\Tc\Tc\hc & $1.018_1$ & $0.1694_1$ & 0.65 \\
$[\text{T}\text{H}\text{T}]_3\text{HT}[\text{T}\text{H}\text{T}]_3$ & 
\Tc\hc\Tc\Tc\hc\Tc\Tc\hc\Tc\hc\Tc\Tc\hc\Tc\Tc\hc\Tc\Tc\hc\Tc & $1.115_9$ & $0.175_5$ & 0.65 \\
$\text{T}\text{H}[\text{T}_{2}\text{H}]_6$ & \Tc\hc\Tc\Tc\hc\Tc\Tc\hc\Tc\Tc\hc\Tc\Tc\hc\Tc\Tc\hc\Tc\Tc\hc & $1.065_4$ & 
$0.181_2$ & 0.65 \\
\midrule
$\text{H}_2\text{T}_3\text{H}\text{T}_2\text{HTH}\text{T}_2\text{HT}\text{H}_2\text{T}_3$ & \hc\hc\Tc\Tc\Tc\hc\Tc\Tc\hc\Tc\hc\Tc\Tc\hc\Tc\hc\hc\Tc\Tc\Tc & $0.95_1$ & $0.145_6$ & 0.6 \\
$[\text{HT}]_2\text{TH}[\text{T}_2\text{H}]_4\text{TH}$ & \hc\Tc\hc\Tc\Tc\hc\Tc\Tc\hc\Tc\Tc\hc\Tc\Tc\hc\Tc\Tc\hc\Tc\hc & $0.808_6$ & $0.142_5$ & 0.6 \\
$\text{H}\text{T}_2\text{H}\text{T}_3\text{H}_2\text{T}_3\text{H}_2\text{T}_2\text{H}\text{T}_2\text{H}$ & \hc\Tc\Tc\hc\Tc\Tc\Tc\hc\hc\Tc\Tc\Tc\hc\hc\Tc\Tc\hc\Tc\Tc\hc & $0.87_1$ & $0.13_2$ & 0.6 \\
$\text{H}[\text{T}_3\text{H}_2]_3\text{T}_3\text{H}$ & \hc\Tc\Tc\Tc\hc\hc\Tc\Tc\Tc\hc\hc\Tc\Tc\Tc\hc\hc\Tc\Tc\Tc\hc & $0.914_4$ & $0.137_2$ & 0.6 \\
$\text{TH}\text{T}_2\text{H}[\text{TH}]_2\text{T}_2\text{H}[\text{TH}]_2\text{T}_2\text{HT}$ & \Tc\hc\Tc\Tc\hc\Tc\hc\Tc\hc\Tc\Tc\hc\Tc\hc\Tc\hc\Tc\Tc\hc\Tc & $0.948_2$ & $0.175_1$ & 0.6 \\
$\text{TH}\text{T}_2\text{H}[\text{T}_2\text{H}_2]_2[\text{T}_2\text{H}]_2\text{T}$ & \Tc\hc\Tc\Tc\hc\Tc\Tc\hc\hc\Tc\Tc\hc\hc\Tc\Tc\hc\Tc\Tc\hc\Tc & $0.967_3$ & $0.161_3$ & 0.6 \\
$\text{TH}\text{T}_2\text{H}\text{T}_2\text{H}_2\text{T}_3\text{H}_2[\text{T}_2\text{H}]_2$ & \Tc\hc\Tc\Tc\hc\Tc\Tc\hc\hc\Tc\Tc\Tc\hc\hc\Tc\Tc\hc\Tc\Tc\hc & $0.918_3$ & $0.152_4$ & 0.6 \\
$\text{HT}[\text{T}_2\text{H}]_2[\text{TH}]_2[\text{T}_2\text{H}]_2\text{TH}$ & \Tc\hc\Tc\Tc\hc\Tc\Tc\hc\Tc\hc\Tc\hc\Tc\Tc\hc\Tc\Tc\hc\Tc\hc & $0.898_2$ & $0.167_1$ & 0.6 \\
$[\text{T}_2\text{H}_2]_2[\text{T}_3\text{H}_2]_2\text{T}_2$ & \Tc\Tc\hc\hc\Tc\Tc\hc\hc\Tc\Tc\Tc\hc\hc\Tc\Tc\Tc\hc\hc\Tc\Tc & $1.011_2$ & $0.157_2$ & 0.6 \\
$\text{T}_2\text{H}_2[\text{T}_3\text{H}_2]_3\text{T}$ & \Tc\Tc\hc\hc\Tc\Tc\Tc\hc\hc\Tc\Tc\Tc\hc\hc\Tc\Tc\Tc\hc\hc\Tc & $0.979_4$ & $0.151_3$ & 0.6 \\
$\text{T}_3\text{H}_3\text{T}_3\text{H}_2\text{T}_3\text{H}_3\text{T}_3$ & \Tc\Tc\Tc\hc\hc\hc\Tc\Tc\Tc\hc\hc\Tc\Tc\Tc\hc\hc\hc\Tc\Tc\Tc & $1.083_3$ & $0.141_2$ & 0.6 \\

	\bottomrule
	\label{tab:crit}
	\end{tabular}
\end{table}

\begin{table}
	\centering
		\caption{Names and architectures of sequences without a critical point. }
\begin{tabular}{ccc} 	
		sequence & architecture    &    	$f_T$\\
		\toprule
$[\text{H}_{2}\text{T}_{3}]_4$ & \hc\hc \Tc\Tc\Tc \hc\hc \Tc\Tc\Tc \hc\hc \Tc\Tc\Tc \hc\hc \Tc\Tc\Tc  & 0.6 \\
$\text{H}_{3}[\text{T}_{2} \text{H}]_4\text{TH}\text{T}_3$  & \hc\hc\hc \Tc\Tc\hc \Tc\Tc\hc \Tc\Tc\hc \Tc\Tc\hc  \Tc\hc\Tc\Tc\Tc & 0.6 \\
$\text{H}_{3}\text{T}_{3} \text{H}_{2}\text{T}_{3}\text{H}\text{T}_{3}\text{H}_{2}\text{T}_{3} $ & \hc\hc\hc\Tc\Tc\Tc \hc\hc \Tc\Tc\Tc \hc \Tc\Tc\Tc \hc\hc \Tc\Tc\Tc   & 0.6 \\
$\text{H}_{3}[\text{T}_4\text{H}]_3\text{H}_{2}$ & \hc\hc\hc\Tc\Tc\Tc\Tc\hc \Tc\Tc\Tc\Tc\hc \Tc\Tc\Tc\Tc\hc \hc\hc & 0.6 \\
$\text{H}_4\text{T}_{12}\text{H}_4$ & \hc\hc\hc\hc\Tc\Tc\Tc\Tc\Tc\Tc\Tc\Tc\Tc\Tc\Tc\Tc\hc\hc\hc\hc& 0.6 \\
$\text{T}_{6}\text{H}_8\text{T}_{6}$& \Tc\Tc\Tc\Tc\Tc\Tc\hc\hc\hc\hc\hc\hc\hc\hc\Tc\Tc\Tc\Tc\Tc\Tc & 0.6 \\
$\text{T}_{12}\text{H}_8$ & \Tc\Tc\Tc\Tc\Tc\Tc\Tc\Tc\Tc\Tc\Tc\Tc\hc\hc\hc\hc\hc\hc\hc\hc & 0.6 \\
	\bottomrule
	\label{tab:nocrit}
	\end{tabular}
\end{table}

\end{document}